\documentclass{vldb}

\usepackage{booktabs} 

\usepackage{graphicx}
\usepackage{amssymb}
\usepackage{algorithm}
\usepackage{stackrel}
\usepackage{algpseudocode}
\usepackage{textcomp}
\PassOptionsToPackage{table}{xcolor}
\usepackage[table]{xcolor}
\usepackage{listings}
\usepackage{subcaption}
\usepackage{amsmath}
\usepackage{mathtools,xparse}
\usepackage{dsfont}
\usepackage{stmaryrd}
\usepackage{paralist}
\usepackage{hyperref}
\hypersetup{hidelinks=true}
\usepackage{cleveref}
\usepackage{lipsum}
\usepackage[normalem]{ulem}
\usepackage{array}
\usepackage{multicol}
\usepackage{enumitem}
\usepackage{etoolbox}
\usepackage{varwidth}
\usepackage{xspace}

\usepackage{multirow}

\makeatletter
\preto{\@verbatim}{\topsep=5pt \parsep=0pt }
\makeatother

\newcolumntype{L}[1]{>{\raggedright\let\newline\\\arraybackslash\hspace{0pt}}m{#1}}
\newcolumntype{C}[1]{>{\centering\let\newline\\\arraybackslash\hspace{0pt}}m{#1}}
\newcolumntype{R}[1]{>{\raggedleft\let\newline\\\arraybackslash\hspace{0pt}}m{#1}}

\newcommand{\comprehension}[2]{\left\{\left.\;{#1}\;\right|\;{#2}\;\right\}}

\newcommand{\tuple}[1]{\left<\;{#1}\;\right>}

\newcommand{\expect}{\mathbb{E}}

\newcommand{\entropy}{\mathcal{H}}

\DeclareMathOperator*{\argmin}{arg\,min}
\newcommand*{\argminl}{\argmin\limits}

\DeclareMathOperator*{\mysum}{\sum}
\newcommand*{\mysuml}{\mysum\limits}

\renewcommand{\vec}[1]{\mathbf{#1}}

\newcommand{\tinysection}[1]{\smallskip \noindent \textbf{#1.}~}

\newcommand{\systemname}{\textsc{LogR}\xspace}
\newcommand{\Errorname}{Reproduction Error\xspace}
\newcommand{\errorname}{\Errorname}
\newcommand{\cqword}[2]{{\footnotesize $\tuple{\text{\lstinline{#2}}, \text{\lstinline{#1}}}$}}
\newcommand{\pattern}{\vec b}
\newcommand{\encoding}{\mathcal E}
\definecolor{light-gray}{gray}{0.75}
\definecolor{mid-gray}{gray}{0.45}

\newcommand{\trimfigurewhitespace}{\vspace*{-5mm}}
\newcommand{\bfcaption}[1]{\caption{\textbf{#1}}}

\begin{document}
\title{Query Log Compression for Workload Analytics}

\toappear{}
\numberofauthors{3}
\author{
\alignauthor
	Ting Xie\\
	\affaddr{University at Buffalo, SUNY}\\
	\email{tingxie@buffalo.edu}
\alignauthor
  Varun Chandola\\
  \affaddr{University at Buffalo, SUNY}\\ 
  \email{chandola@buffalo.edu}
\alignauthor
	Oliver Kennedy\\
	\affaddr{University at Buffalo, SUNY}\\ 
	\email{okennedy@buffalo.edu}
}
\newtheorem{example}{Example}
\newtheorem{proposition}{Proposition}
\newtheorem{lemma}{Lemma}


\maketitle

\begin{abstract}
Analyzing database access logs is a key part of performance tuning, intrusion detection, benchmark development, and many other database administration tasks.
Unfortunately, it is common for production databases to deal with millions or even more queries each day, so these logs must be summarized before they can be used.
Designing an appropriate summary encoding requires trading off between conciseness and information content.
For example: simple workload sampling may miss rare, but high impact queries.
In this paper, we present \systemname, a lossy log compression scheme suitable use for many automated log analytics tools, as well as for human inspection.
We formalize and analyze the space/fidelity trade-off in the context of a broader family of ``pattern'' and ``pattern mixture'' log encodings to which \systemname belongs.
We show through a series of experiments that \systemname compressed encodings can be created efficiently, come with provable information-theoretic bounds on their accuracy, and outperform state-of-art log summarization strategies.

\end{abstract}

\section{Introduction}
Automated analysis of database access logs is critical for solving a wide range of problems, from database performance tuning~\cite{Bruno:2005:APD:1066157.1066184}, to compliance validation~\cite{Dwork2006} and query recommendation~\cite{chatzopoulou2011querie}. 
For example, the Peloton self-tuning database~\cite{DBLP:conf/cidr/PavloAALLMMMPQS17} searches for optimal configurations by repeatedly simulating database performance based on statistical properties of historical queries.
Unfortunately, query logs for production databases can grow to be large ---
A recent study of queries at a major US bank for a period of 19 hours found nearly 17 million SQL queries and over 60 million stored procedure executions~\cite{Kul:2016:EAQ:2872518.2888608} --- and computing these properties from the log itself is slow.

Tracking only a sample of these queries is not sufficient, as rare queries can disproportionately affect database performance, for example, if they benefit from an otherwise unnecessary index.
Rather, we need a compressed \emph{summary} of the log on which we can compute aggregate statistical properties.
The problems of compression and summarization have been studied extensively (e.g.,~\cite{1055714,Ziv:2006:CIS:2263333.2269400,huffman1952method,eckart1936approximation,Blei:2012:PTM:2133806.2133826,Wang:2009:MSU:1667583.1667675,Knight:2002:SBS:604203.604207}). 
However, these schemes either require the use of heavyweight inference to desired statistical measures, or produce unnecessarily large encodings.

In this paper, we adapt ideas from pattern mining and summarization~\cite{Mampaey:2012:SDS:2382577.2382580,ElGebaly:2014:IIE:2735461.2735467} to propose a middle-ground: \systemname, a summarization scheme that facilitates efficient (both in terms of storage and time) approximation of workload statistics.
By adjusting a tunable parameter in \systemname, users can choose to obtain a high-fidelity, albeit large summary, or obtain a more compact summary with lower fidelity.
Constructing the summary that best balances compactness and fidelity is challenging, as the search space of candidate summaries is combinatorially large~\cite{Mampaey:2012:SDS:2382577.2382580,ElGebaly:2014:IIE:2735461.2735467}.
\systemname offers a new approach to summary construction that avoids searching this space, making inexpensive, accurate computation of aggregate workload statistics possible.
As a secondary benefit, the resulting summaries are also human-interpretable.

\systemname does not admit closed-form solutions to classical fidelity measures like information loss, so we propose an alternative called \emph{\errorname}.
We show through a combination of analytical and experimental evidence that \errorname is highly correlated with several classical measures of encoding fidelity.

\systemname-compressed data relies on a codebook based on structural elements like \texttt{SELECT} items, \texttt{FROM} tables, or conjunctive \texttt{WHERE} clauses~\cite{Aligon2014}.
This codebook provides a bi-directional mapping from SQL queries to a bit-vector encoding and back again, reducing the compression problem to one of compactly encoding a collection of feature-vectors.
We further simplify the problem by observing that a common theme in use cases like automated performance tuning or query recommendation is the need for predominantly aggregate workload statistics.
As these are order-independent, we are able to focus exclusively on compactly representing \emph{bags} of feature-vectors.

\systemname works by identifying groups of co-occurring structural elements that we call patterns.  
We define a family of \emph{pattern encodings} of access logs, which map patterns to their frequencies in the log.
For pattern encodings, we consider two idealized measures of fidelity: 
(1) Ambiguity, which measures how much room the encoding leaves for interpretation; and 
(2) Deviation, which measures how reliably the encoding approximates the original log.
Neither Ambiguity nor Deviation can be computed efficiently for pattern encodings.
Hence we propose a measure called \emph{\errorname} that is efficiently computable and that closely tracks both Ambiguity and Deviation.

In general, the size of the encoding is inversely related with \errorname: The more detailed the encoding, the more faithfully it represents the original log.
Thus, log compression may be defined as a search over the space of pattern based encodings to identify the one that best trades off between these two properties.
Unfortunately, searching for such an ideal encoding from the space can be computationally expensive~\cite{ElGebaly:2014:IIE:2735461.2735467,Mampaey:2012:SDS:2382577.2382580}.
To overcome this limitation, we reduce the search space by first clustering entries in the log and then encoding each cluster separately, an approach that we call \textit{pattern mixture encoding}.
Finally we identify a simple approach to encoding individual clusters that we call \textit{naive mixture encodings}, and show experimentally that it produces results competitive with more powerful techniques for log compression and summarization.

Concretely, in this paper we make the following contributions:
(1) We define two families of compression for query logs: pattern and pattern mixture, 
(2) We define a computationally efficient measure, \errorname, and demonstrate that it is a close approximation of Ambiguity and Deviation (two commonly used measures),
(3) We propose a clustering-based approach to efficiently search for naive mixture encodings, and show how these encodings can be further optimized, and, 
(4) We experimentally validate \systemname and show that it produces more precise encodings faster than several state-of-the-art pattern encoding algorithms.

\tinysection{Roadmap}
The paper is organized as follows: Section~\ref{sec:problemdefinition} formally defines the log compression problem and the summary representation;
Section~\ref{sec:analyzingsummaries} then defines information loss for the summaries;
Section~\ref{sec:practicalrepresentativeness} explains the difficulty in computing the theoretical loss measure and provides a practical alternative;
Section~\ref{sec:patternmixtureencodings} motivates data partitioning and generalizes the practical loss measure to partitioned data;
Section~\ref{sec:constructingencodings} then introduces the proposed \systemname compression scheme;
Section~\ref{sec:experiments} empirically validates the practical loss measure and evaluates the effectiveness of \systemname compression by comparing it with two state-of-the-art summarization methods;
Section~\ref{sec:evaluatingalternativeapplications} empirically verifies the effectiveness of \systemname by evaluating it under the applications of the two comparison methods;
Section~\ref{sec:backgroundandrelatedwork} discusses related work and Section~\ref{sec:conclusion} concludes the paper.


\section{Problem Definition}
\label{sec:problemdefinition}
In this section, we introduce and formally define the log compression problem.
We begin by exploring several applications that need to repeatedly analyze query logs.

\tinysection{Index Selection}
Selecting an appropriate set of indexes requires trading off between update costs, access costs, and limitations on available storage space.
Existing strategies for selecting a (near-)optimal set of indexes typically repeatedly simulate database performance under different combinations of indexes, which in turn requires repeatedly estimating the frequency with which specific predicates appear in the workload.
For example, if \lstinline{status = ?} occurs in $90\%$ of the queries in a workload, a hash index on \lstinline{status} is beneficial.

\tinysection{Materialized View Selection}
The results of joins or highly selective selection predicates are good candidates for materialization when they appear frequently in the workload.  
Like index selection, view selection is a non-convex optimization problem, typically requiring exploration by repeated simulation, which in turn requires repeated frequency estimation over the workload.

\tinysection{Online Database Monitoring}
In production settings, it is common to monitor databases for atypical usage patterns that could indicate a serious bug or security threat.
When query logs are monitored, it is often done retrospectively, some hours after-the-fact~\cite{Kul:2016:EAQ:2872518.2888608}.  
To support real-time monitoring it is necessary to quickly compute the frequency of a particular class of query in the system's typical workload.

\smallskip

In each case, the application's interactions with the log amount to counting queries that have specific features: selection predicates, joins, or similar.

\subsection{Preliminaries and Notation}
\label{sec:notation}
Let $L$ be a log, or a finite collection of queries $\vec q \in L$.
We write $f \in \vec q$ to indicate that $\vec q$ has some \emph{feature} $f$, such as a specific predicate or table in its \lstinline{FROM} clause.
We assume (1) that the universe of features in both a log and a query is enumerable and finite, (2) that the features are selected to suit specific applications and (3) optionally that a query is isomorphic to its feature set (motivated in Section~\ref{sec:communicatinginformationcontent}).  
We outline one approach to extracting features that satisfies all three assumptions below.
We abuse syntax and write $\vec q$ to denote both the query itself, as well as the set of its features.  

Let $\pattern$ denote some set of features $f \in \pattern$, which we call a \emph{pattern}.
We write these sets using vector notation: $\pattern=(x_1,\ldots,x_n)$ where $n$ is the number of distinct features appearing in the entire log and $x_i$ indicates the presence (absence) of $i$th feature with a 1 (resp., 0).  
For any two patterns $\pattern$, $\pattern'$, we say that $\pattern'$ \emph{is contained} in $\pattern$ if $\pattern' \subseteq \pattern$.  Equivalently, with $\pattern=(x_1,\ldots,x_n)$ and $\pattern'=(x_1',\ldots,x_n')$:
$$\pattern' \subseteq \pattern\;\;\; \equiv \;\;\;\forall i,\, x'_i\leq x_i$$
Our goal then is to be able to query logs for the number of times a pattern $\pattern$ appears,
$|\comprehension{\vec q}{\vec q \in L \wedge \pattern \subseteq \vec q}|$

\subsection{Coding Queries}

For this paper, we specifically adopt the feature-extraction conventions of a query summarization scheme by Aligon et al.~\cite{Aligon2014}.
In this scheme, each feature is one of the following three query elements:
(1) a table or sub-query in the \texttt{FROM} clause,
(2) a column in the \texttt{SELECT} clause, and 
(3) a conjunctive atom of the \texttt{WHERE} clause.
\begin{example}
\label{exampleQuery}
Consider the following example query.
\begin{lstlisting}
SELECT _id, sms_type, _time FROM Messages
WHERE status=? AND transport_type=?
\end{lstlisting}

\noindent This query uses 6 features: 
\cqword{SELECT}{sms\_type},
\cqword{SELECT}{\_id},
\cqword{SELECT}{\_time},
\cqword{FROM}{Messages}, 
\cqword{WHERE}{status=?}, \linebreak and
\cqword{WHERE}{transport\_type=?}
\end{example}

Although this scheme is simple and limited to conjunctive queries (or queries with a conjunctive equivalent), it fulfills all three assumptions we make on feature extraction schemes.  
The features of a query (and consequently a log) are enumerable and finite, and the feature set of the query is isomorphic (modulo commutativity and column order) to the original query.
Furthermore, even if a query is not itself conjunctive, it often has a conjunctive equivalent.
We quantify this statement with Table~\ref{table:datasummary}, which provides two relevant data points from production query logs; In both cases, \emph{all} logged queries can be rewritten into equivalent queries compatible with the Aligon scheme.


Although we do not explore more advanced feature encoding schemes in detail here, we direct the interested reader to work on query summarization~\cite{makiyama2015text,aouiche2006,Kul:2016:EAQ:2872518.2888608}.
For example, a scheme by Makiyama et. al.~\cite{makiyama2015text} also captures aggregation-related features like group-by columns, while an approach by Kul et. al.~\cite{Kul:2016:EAQ:2872518.2888608} encodes partial tree-structures in the query.

\subsection{Log Compression}
\label{queryrepresentation}

As a lossy form of compression, \systemname only approximates the information content of a query log.
We next develop a simplified form of \systemname that we call pattern-based encoding, and develop a framework for reasoning about the fidelity of a \systemname-compressed log.
As a basis for this framework, we first reframe the information content of a query log to allow us to adapt classical information-theoretical measures of information content.

\subsubsection{Information Content of Logs}
\label{sec:informationcontentoflogs}

We define the information content of the log as a distribution $p(Q\;|\;L)$ of queries $Q$ drawn uniformly from the log.

\begin{example}
\label{distributionExample}
Consider the following query log, which consists of four conjunctive queries.
\begin{lstlisting}
1. SELECT _id FROM Messages WHERE status = ?
2. SELECT _time FROM Messages 
          WHERE status = ? AND sms_type = ?
3. SELECT _id FROM Messages WHERE status = ?
4. SELECT sms_type, _time FROM Messages 
          WHERE sms_type = ?
\end{lstlisting}
Drawing uniformly from the log, each entry will appear with probability $\frac{1}{4} = 0.25$.
The query $q_1$ ($=q_3$) occurs twice, so the probability of drawing it is double that of the others (i.e., $p(\vec{q}_1\;|\;L) = p(\vec{q}_3\;|\;L) = \frac{2}{4} = 0.5$)
\end{example}

Treating a query as a vector of its component features, we can define a query $\vec{q}=(x_1,\ldots,x_n)$ to be an observation of the multivariate distribution over variables $Q = (X_1,\ldots,X_n)$ corresponding to features.
The event $X_i = 1$ occurs if feature $i$ appears in a uniformly drawn query.

\begin{example}
Continuing, the universe of features for this query log is (1)~\cqword{SELECT}{\_id}, (2)~\cqword{SELECT}{\_time},\linebreak (3)~\cqword{SELECT}{sms\_type}, (4)~\cqword{WHERE}{status = ?}, \linebreak (5)~\cqword{WHERE}{sms\_type = ?}, and (6)~\cqword{FROM}{Messages}.
Accordingly, the queries can be encoded as feature vectors, with fields counting each feature's occurrences:
{\small
$\vec{q}_1 = \tuple{1, 0, 0, 1, 0, 1}$, 
$\vec{q}_2 = \tuple{0, 1, 0, 1, 1, 1}$, 
$\vec{q}_3 = \tuple{1, 0, 0, 1, 0, 1}$, 
$\vec{q}_4 = \tuple{0, 1, 1, 0, 1, 1}$
}
\end{example}

\begin{figure}
 \centering
\begin{subfigure}{\columnwidth}
  {\small
    \begin{tabular}{rp{60mm}}
    \textbf{SELECT} & 
        \texttt{sms\_type},
        \textcolor{light-gray}{\texttt{external\_ids}},
        \texttt{\_time},
        \texttt{\_id}\\ 
    \textbf{FROM} &
        \texttt{messages}\\ 
    \textbf{WHERE} &
        \textcolor{mid-gray} {\texttt{(sms\_type=?)}} $\wedge$
        \texttt{(status=?)}   
    \end{tabular}
  }
  \caption{\textit{Correlation-ignorant}: Features are highlighted independently}
  \label{fig:screenshots:nocorrelation}
\end{subfigure}\\[2mm]
\begin{subfigure}{\columnwidth}
  {\centering
    \fbox{\parbox{0.96\textwidth}{
      {\small \centering
        \textbf{SELECT} \texttt{sms\_type} \textbf{FROM} \texttt{messages} \textbf{WHERE} \texttt{sms\_type=?}
      }\\
      \textcolor{mid-gray}{\small \centering
        \textbf{SELECT} \texttt{sms\_type} \textbf{FROM} \texttt{messages} \textbf{WHERE} \texttt{status=?} 
      }
    }}
  }
  \caption{\textit{Correlation-aware}: Pattern groups are highlighted together.
  }
  \label{fig:screenshots:correlation}  
\end{subfigure}\\[2mm]
\bfcaption{Example Encoding Visualizations}
\label{fig:screenshots}
\trimfigurewhitespace
\end{figure} 


\tinysection{Patterns}
Our target applications require us to count the number of times features (co-)occur in a query.  
For example, materialized view selection requires counting tables used together in queries.
Motivated by this observation, we begin by defining a broad class of \emph{pattern based encodings} that directly encode co-occurrence probabilities.
A \emph{pattern} is an arbitrary set of features $\pattern=(x_1,\ldots,x_n)$ that may co-occur together.
Each pattern captures a piece of information from the distribution $p(Q\;|\;L)$.
In particular, we are interested in the probability of uniformly drawing a query $\vec{q}$ from the log that \textit{contains} the pattern $\vec b$ (i.e., $\vec{q}\supseteq\pattern$): \vspace*{-1mm}
\begin{center}
{\small $p(Q\supseteq\pattern\;|\;L)=\sum_{\vec{q}\in L\land \vec{q}\supseteq \pattern}p(\vec{q}\;|\;L)$}
\end{center}\vspace*{-1mm}
\noindent When it is clear from context, we abuse notation and write $p(\cdot)$ instead of $p(\cdot\;|\;L)$.
Recall that $p(Q)$ can be represented as a joint distribution over variables $(X_1,\ldots,X_n)$ and probability $p(Q\supseteq\pattern)$ is thus equivalent to the marginal probability $p(X_1\geq x_1,\ldots,X_n\geq x_n)$ of pattern $\vec b$.

\tinysection{Pattern-Based Encodings}
Denote by $\encoding_{max} : \{0,1\}^n \rightarrow [0,1]$, the mapping from the space of all possible patterns $\pattern\in\{0,1\}^n$ to their marginals. 
A \emph{pattern based encoding} $\encoding$ is any such partial mapping $\encoding \subseteq \encoding_{max}$. 
We denote the marginal of pattern $\pattern$ in encoding $\encoding_L$ by $\encoding_L[\pattern] $ ($= p(Q\supseteq\pattern\;|\;L)$).
When it is clear from context, we abuse syntax and also use $\encoding$ to denote the set of patterns it maps (i.e., $domain(\encoding)$).
Hence, $|\encoding|$ is the number of mapped patterns, which we call the encoding's \emph{Verbosity}.
A \emph{pattern based encoder} is any algorithm $\texttt{encode}(L, \epsilon) \mapsto \encoding$ whose input is a log $L$ and whose output is a set of patterns $\encoding$, with Verbosity thresholded at some integer $\epsilon$.
Many pattern mining algorithms~\cite{ElGebaly:2014:IIE:2735461.2735467,Mampaey:2012:SDS:2382577.2382580} can be used for this purpose.

\subsubsection{Communicating Information Content}
\label{sec:communicatinginformationcontent}
A side-benefit of pattern based encodings is that, under the assumption of isomorphism in Section~\ref{sec:notation}, patterns can be translated to their query representations and used for human analysis of the log.
Figure~\ref{fig:screenshots} shows two examples.
The approach illustrated in Figure~\ref{fig:screenshots:nocorrelation} uses shading to show each feature's frequency in the log, and communicates frequently occurring constraints or attributes.
This approach might, for example, help a human to manually select indexes.
A second approach illustrated in Figure~\ref{fig:screenshots:correlation} conveys correlations, showing the frequency of entire patterns.
Appendix~\ref{appendix:naivemixturesummaryvisualization} explores interpretable visualizations of pattern based summaries in greater depth.

\section{Information Loss}
\label{sec:analyzingsummaries}


Our goal is to encode the distribution $p(Q)$ as a set of patterns: obtaining a less verbose encoding (i.e., with fewer patterns), while also ensuring that the encoding captures $p(Q)$ with minimal information loss.
In this section, we defines information loss for pattern based encodings.

\subsection{Lossless Summaries}
\label{sec:representativeness:idealdef}
To establish a baseline for measuring information loss, we begin with the extreme cases.
At one extreme, an empty encoding ($|\encoding| = 0$) conveys no information.
At the other extreme, we have the encoding $\encoding_{max}$ which is the full mapping from all patterns. 
Having this encoding is a sufficient condition to exactly reconstruct the original distribution $p(Q)$. 
\vspace*{-4mm}
\begin{proposition}
\label{PROPOSITION:LOSSLESSSUMMARY} 
For any query $\vec{q}=(x_1,\ldots,x_n)\in\mathbb{N}^n$,
the probability of drawing exactly $\vec{q}$ at random from the log (i.e., $p(X_1=x_1,\ldots,X_n=x_n)$) is computable, given $\encoding_{max}$.
\end{proposition}
See Appendix~\ref{appendix:losslesssummary} for proof of the proposition.


\subsection{Lossy Summaries}
\label{sec:lossysummaries}
Although lossless, $\encoding_{max}$ is also verbose. 
Hence, we will focus on lossy encodings that can be less verbose.
A lossy encoding $\encoding \subset \encoding_{max}$ may not be able to precisely identify the distribution $p(Q)$, but can still be used to approximate it.
We characterize the information content of a lossy encoding $\encoding$ by defining a \emph{space} (denoted by $\Omega_\encoding$) of distributions $\rho \in \Omega_\encoding$ allowed by an encoding $\encoding$.
This space is defined by constraints as follows:
First, we have the general properties of probability distributions:
\begin{center}
$\forall \vec{q}\in\mathbb{N}^n:\rho(\vec{q})\geq 0$
\hspace{10mm}
$\sum_{\vec{q}\in\mathbb{N}^n}\rho(\vec{q})=1$
\end{center}
Each pattern $\vec b$ in the encoding $\encoding$ constrains the marginal probability over its component features:
\begin{equation*}
\forall \vec{b} \in domain(\encoding)  :\;\; \encoding[\vec{b}] = \sum\nolimits_{\vec{q}\supseteq\vec{b}} \rho(\vec{q}) \;\;\;
\end{equation*}
Note that the dual constraints $1-\encoding[\vec{b}]=\sum_{\vec{q}\not\supseteq\vec{b}} \rho(\vec{q})$ are redundant under constraint $\sum_{\vec{q}\in\mathbb{N}^n}\rho(\vec{q})=1$.

The resulting space $\Omega_\encoding$ is the set of all query logs, or equivalently the set of all possible distributions of queries, that obey these constraints.
From the outside observer's perspective, the distribution $\rho\in\Omega_\encoding$ that the encoding conveys is ambiguous: We model this ambiguity with a random variable $\mathcal P_\encoding$ with support $\Omega_\encoding$.
The true distribution $p(Q)$ derived from the query log must appear in $\Omega_\encoding$, denoted $p(Q)\equiv\rho^*\in\Omega_\encoding$ (i.e., $p(\mathcal P_\encoding = \rho^*) > 0$). 
Of the remaining distributions $\rho$ admitted by $\Omega_\encoding$, it is possible that some are more likely than others.
For example, a query containing a column (e.g., \texttt{status}) is only valid if it also references a table that contains the column (e.g., \texttt{Messages}).
This prior knowledge may be modeled as a prior on the distribution of $\mathcal P_\encoding$ or by an additional constraint.
However, for the purposes of this paper, we take the uninformed prior by assuming that $\mathcal P_\encoding$ is uniformly distributed over $\Omega_\encoding$:
\begin{equation*}
\label{uniformprior}
p(\mathcal P_\encoding = \rho) = 
\begin{cases}
\frac{1}{|\Omega_\encoding|} & \text{if } \rho \in \Omega_\encoding\\
0 & \text{otherwise}
\end{cases}
\end{equation*}


\tinysection{Naive Encodings}
One specific family of lossy encodings that treats each feature as being independent (e.g., as in Figure~\ref{fig:screenshots:nocorrelation}) is of particular interest to us.  
We call this family \emph{naive encodings}, and return to it throughout the rest of the paper.
A naive encoding is composed of all patterns that have exactly one feature with non-zero marginal.
$$\comprehension{\vec b = (0, \ldots, 0, x_i, 0, \ldots, 0)}{i \in [1,n],\; x_i = 1}$$



\subsection{Idealized Information Loss Measures}
\label{sec:idealizedrepresentativenessmeasures}
Based on the space of distributions constrained by the encoding, the information loss of an encoding can be considered from two related, but subtly distinct perspectives:
(1) \emph{Ambiguity} measures how much room the encoding leaves for interpretation, 
(2) \emph{Deviation} measures how reliably the encoding approximates the target distribution $p(Q)$.

\smallskip
\tinysection{Ambiguity}
We define the Ambiguity $\text{I}(\encoding)$ of an encoding as the entropy of the random variable $\mathcal P_\encoding$. 
The higher the entropy, the less precisely $\encoding$ identifies a specific distribution.
$$\text{I}(\encoding) = \mysuml_{\rho}p(\mathcal P_\encoding = \rho)\log \left(p(\mathcal P_\encoding=\rho)\right)$$

\tinysection{Deviation}
The deviation from any permitted distribution $\rho$ to the true distribution $\rho^*$ can be measured by the Kullback-Leibler (K-L) divergence~\cite{kullback1951} (denoted $\mathcal{D}_{KL}(\rho^*||\rho)$).
We define the Deviation $\text{d}(\encoding)$ of a encoding as the expectation of the K-L divergence over all permitted $\rho \in \Omega_\encoding$:
$$\text{d}(\encoding)=\expect_{\mathcal{P}_\encoding}\left[\mathcal{D}_{KL}(\rho^*||\mathcal{P}_\encoding)\right] = \sum_{\rho \in \Omega_\encoding} p(\mathcal P_\encoding = \rho) \cdot \mathcal{D}_{KL}(\rho^*||\rho)$$



\tinysection{Limitations}
There are two limitations to these idealized measures in practice.
First, K-L divergence is not defined from any probability measure $\rho^*$ that is not \emph{absolutely continuous} with respect to a second (denoted  $\rho^*\ll\rho$). 
Second, neither Deviation nor Ambiguity has a closed-form formula. 

\section{Practical Loss Measure}
\label{sec:practicalrepresentativeness}
Computing either Ambiguity or Deviation requires enumerating the entire space of possible distributions, or an approximation.
One approach to estimating either measure is repeatedly sampling from, rather than enumerating the space.  
However, accurate measures require a large number of samples, rendering this approach similarly infeasible.
In this section, we propose a faster approach to assessing the fidelity of a pattern encoding.
Specifically, we select a single representative distribution $\overline \rho_\encoding$ from the space $\Omega_\encoding$, and use $\overline \rho_\encoding$ to approximate both Ambiguity and Deviation.


\subsection{\Errorname}
\label{sec:maximumentropydistribution}

\tinysection{Maximum Entropy Distribution}
The representative distribution is chosen by applying maximum entropy principle~\cite{jaynes2003probability} which is commonly used in pattern-based summarization~\cite{ElGebaly:2014:IIE:2735461.2735467,Mampaey:2012:SDS:2382577.2382580}.
That is, we select the distribution $\overline{\rho}_\encoding$ with maximum entropy:
$$\overline{\rho}_\encoding=\argminl_{\rho\in\Omega_\encoding}-\entropy(\rho)\;\;\;\;\;\;\text{where }\entropy(\rho)=\mysuml_{\vec{q}\in\mathbb{N}^n}-\rho(\vec{q})\log \rho(\vec{q})$$
The maximum entropy distribution $\overline{\rho}_\encoding$ best represents the current state of knowledge.
That is, a distribution with lower entropy assumes additional constraints derived from patterns that we do not know and one with higher entropy violates the constraints from patterns we do know.

Maximizing an objective function belonging to the exponential family (entropy in our case) under a mixture of linear equalities/inequality constraints is a convex optimization problem~\cite{Boyd:2004:CO:993483} which guarantees a \emph{unique} solution and can be efficiently solved~\cite{darroch1972}, using the cvx toolkit~\cite{cvx}\cite{ODonoghue2016}, and/or by \textit{iterative scaling}~\cite{ElGebaly:2014:IIE:2735461.2735467,Mampaey:2012:SDS:2382577.2382580}.
For naive encodings specifically, we can assume independence between each feature $X_i$.  
Under this assumption, $\overline{\rho}_\encoding$ has a closed-form solution:
\begin{equation}
\label{eqn:naiveencodingformula}
\overline{\rho}_\encoding(\vec{q})=\prod_{i}p(X_i=x_i)\;\;\;\;\;\;\text{where } \vec{q}=(x_1,\ldots,x_n)
\end{equation} 
%
Using~\eqref{eqn:naiveencodingformula}, 
we define \textit{\errorname} $e(\encoding)$ as the entropy difference between the representative and true distributions:
$$e(\encoding)=\entropy(\overline{\rho}_\encoding)-\entropy(\rho^*)\;\;\;\;\;\text{where }\overline{\rho}_\encoding=\argminl_{\rho\in\Omega_\encoding}-\entropy(\rho)$$

\subsection{Practical vs Idealized Information Loss}
\label{sec:validateencodingerror}

In this section we prove that \errorname closely parallels Ambiguity. 
We define a partial order lattice over encodings and show that for any pair of encodings on which the partial order is defined, a like relationship is implied for both \errorname and Ambiguity.
We supplement the proofs given in this section with an empirical analysis relating \errorname to Deviation in Section~\ref{sec:motivateencodingerror}.


\tinysection{Containment}
We define a partial order over encodings $\leq_\Omega$ based on \textit{containment} of their induced spaces $\Omega_\encoding$:
$$\encoding_1 \leq_\Omega \encoding_2\;\;\;\equiv\;\;\;\Omega_{\encoding_1} \subseteq \Omega_{\encoding_2}$$
That is, one encoding (i.e., $\encoding_1$) precedes another (i.e., $\encoding_2$) when all distributions admitted by the former encoding are also admitted by the latter.


\tinysection{Containment Captures \Errorname}
We first prove that the total order given by \errorname is a superset of the partial order $\leq_\Omega$. 
\begin{lemma}
\label{prop:monotone}
For any two encodings $\encoding_1,\encoding_2$ with induced spaces $\Omega_{\encoding_1},\Omega_{\encoding_2}$ and maximum entropy distributions $\overline{\rho}_{\encoding_1},\overline{\rho}_{\encoding_2}$ it holds that $\encoding_1 \leq_\Omega \encoding_2\to e(\encoding_1)\leq e(\encoding_2)$. 
\end{lemma}
\begin{proof} 
Firstly $\Omega_{\encoding_2}\supseteq \Omega_{\encoding_1}\to\overline{\rho}_{\encoding_1}\in\Omega_{\encoding_2}$. Since $\overline{\rho}_{\encoding_2}$ has the maximum entropy among all distributions $\rho\in\Omega_{\encoding_2}$, we have $\entropy(\overline{\rho}_{\encoding_1})\leq\entropy(\overline{\rho}_{\encoding_2})\equiv e(\encoding_1)\leq e(\encoding_2)$. 
\end{proof}


\tinysection{Containment Captures Ambiguity} Next, we show that the partial order based on containment implies a like relationship between Ambiguities of pairs of encodings.
\begin{lemma}
\label{prop:containmentrepresentativeness}
Given encodings $\encoding_1,\encoding_2$ with uninformed prior on $\mathcal{P}_{\encoding_1},\mathcal{P}_{\encoding_2}$, it holds that $\encoding_1 \leq_\Omega \encoding_2\to \text{I}(\encoding_1)\leq \text{I}(\encoding_2)$.
\end{lemma}
\begin{proof}
Given an uninformed prior: $\text{I}(\encoding)=\log|\Omega_\encoding|$. Hence $\encoding_1 \leq_\Omega \encoding_2\to|\Omega_{\encoding_1}|\leq|\Omega_{\encoding_2}|\to \text{I}(\encoding_1)\leq \text{I}(\encoding_2)$ 
\end{proof}

\section{Pattern Mixture Encodings}
\label{sec:patternmixtureencodings}
Thus far we have defined the problem of log compression, treating the query log as a single joint distribution $p(Q)$ that captures the frequency of feature occurrence and/or co-occurrence.
Patterns capture positive information about correlations.
However in cases like logs of \textit{mixed} workloads, there are also many cases of anti-correlation between features.
For example, consider a log that includes queries drawn from two workloads with disjoint feature sets.
Pattern based summaries can not convey such anti-correlations easily.
As a result, while patterns including features from both workloads never actually occur in the log, a pattern-based summary of the log will indicate a non-zero marginal.
Identifying significant workload variation, as might be caused by misuse or malicious workload-injection (mixture), is relevant to intrusion detection systems~\cite{DBLP:conf/trustcom/KulUC18}.
In addition, capturing anti-correlations helps to reduce data dimensionality and improves both the runtime and accuracy of state-of-the-art pattern mining algorithms (See Section~\ref{sec:classicallaserlightandmtv} and~\ref{sec:generalizinglaserlightandmtv}).

In this section, we propose a generalization of pattern encodings where the log is modeled not as a single probability distribution, but rather as a mixture of several simpler distributions.
The resulting encoding is likewise a mixture: Each component of the mixture of distributions is stored independently.
Hence, we refer to it as a \emph{pattern mixture encoding}, and it forms the basis of \systemname compression.

We first focus on a simplified form of this problem, where we only mix \emph{naive} pattern encodings (we explore more general mixtures in Section~\ref{sec:naivemixtureencodingrefinement}).
We first refer to the resulting scheme as \textit{naive mixture encodings}, and give examples of the encoding, as well as potential visualizations in Section~\ref{sec:naivemixtureencoding}.
Then we generalize \errorname and Verbosity for pattern mixture encodings in Section~\ref{sec:generalizedinformationlossmeasures}.
Finally, with generalized encoding evaluation measures, we evaluate several encoding strategies based on different clustering methods for creating naive mixture encodings.

\subsection{Example: Naive Mixture Encodings}
\label{sec:naivemixtureencoding}
\noindent Consider a toy query log with only 3 conjunctive queries. \\[-5mm]
\begin{enumerate}
\item \lstinline{SELECT id FROM Messages WHERE status = ?}\\[-6mm]
\item \lstinline{SELECT id FROM Messages}\\[-6mm]
\item \lstinline{SELECT sms_type FROM Messages}\\[-5mm]
\end{enumerate}
The vocabulary of this log consists of 4 features:
\cqword{SELECT}{id},
\cqword{SELECT}{sms\_type},
\cqword{FROM}{Messages}, and
\cqword{WHERE}{status = ?}.
Re-encoding the three queries as vectors, we get: 
$$\text{1.}\tuple{1, 0, 1, 1} \hspace{10mm} \text{2.} \tuple{1, 0, 1, 0} \hspace{10mm} \text{3.} \tuple{0, 1, 1, 0}$$
A naive encoding of this log 
can be expressed as:
$$\tuple{\hspace{1mm}\frac{2}{3}, \hspace{2mm} \frac{1}{3}, \hspace{2mm} 1, \hspace{2mm} \frac{1}{3}\hspace{1mm}}$$
This encoding captures that all queries in the log pertain to the Messages table, but obscures the relationship between the remaining features.
For example, this encoding obscures the anti-correlation between \lstinline{id} and \lstinline{sms_type}.
Similarly, the encoding hides the association between \lstinline{status = ?} and \lstinline{id}.  
Such relationships are critical for evaluating the effectiveness of views or indexes.

\begin{example}
\label{naivemixtureencodingexample}
The maximal entropy distribution for a naive encoding assumes that features are independent.
Assuming independence, the probability of query 1 from the log is: 
$$
 p(\texttt{\small id})\cdot
 p(\neg \texttt{\small sms\_type})\cdot
 p(\texttt{\small Messages})\cdot
 p(\texttt{\small status=?}) =
  \frac{4}{27} \approx 0.148
$$
This is a significant difference from the true probability of this query (i.e., $\frac{1}{3}$).  
Conversely queries not in the log, such as the following, have non-zero probability in the encoding. 
\begin{lstlisting}
 SELECT sms_type FROM Messages WHERE status = ?
\end{lstlisting}
\vspace*{-3mm}
$$
 p(\neg\texttt{\small id})\cdot
 p(\texttt{\small sms\_type})\cdot
 p(\texttt{\small Messages})\cdot
 p(\texttt{\small status=?}) =
  \frac{1}{27} \approx 0.037
$$
\end{example}

To achieve a more faithful representation of the original log, we could partition it into two components, with the corresponding encoding parameters:
\begin{center}
\begin{tabular}{cp{6mm}c}
\underline{\textbf{Partition 1} ($L_1$)} && \underline{\textbf{Partition 2} ($L_2$)} \\[1.5mm]
$(1, 0, 1, 1)$ \hspace{3mm} $(1, 0, 1, 0)$ &&  $(0, 1, 1, 0)$\\
$\downarrow$\hspace{8mm}$\downarrow$ && $\downarrow$\\
$\tuple{\hspace{1mm}1, \hspace{2mm} 0, \hspace{2mm} 1, \hspace{2mm} \frac{1}{2}\hspace{1mm}}$ && 
$\tuple{\hspace{1mm}0, \hspace{2mm} 1, \hspace{2mm} 1, \hspace{2mm} 0\hspace{1mm}}$
\end{tabular}
\end{center}

The resulting encoding only has one non-integral probability: $p(\text{\lstinline{status = ?}}\;|\;L_1) = 0.5$.  
Although there are now two encodings, the encodings are not ambiguous. The feature \lstinline{status = ?} appears in exactly half of the log entries, and is indeed independent of the other features.  
All other attributes in each encoding appear in all queries in their respective partitions.  
Furthermore, the maximum entropy distribution induced by each encoding is exactly the distribution of queries in the compressed log.  
Hence, the \errorname is zero for both of the two encodings.

\subsection{Generalized Encoding Fidelity}
\label{sec:generalizedinformationlossmeasures}
We next generalize our definitions of \errorname and Verbosity from pattern to pattern mixture encodings.
Suppose query log $L$ has been partitioned into $K$ clusters with $L_i$, $S_i$, $\overline{\rho}_{S_i}$ and $\rho^*_i$ (where $i \in [1,K]$) representing the log of queries, encoding, maximum entropy distribution, and true distribution (respectively) for $i$th cluster. 
First, observe that the distribution for the whole log (i.e., $\rho^*$) is the sum of distributions for each partition (i.e., $\rho^*_i$) weighted by proportion of queries (i.e., $\frac{|L_i|}{|L|}$) in the partition.
$$\rho^*(\vec{q})=\mysuml_{i=1,\ldots,K}w_i \cdot \rho^*_i(\vec{q}) \hspace{10mm}\text{where } w_i=\frac{|L_i|}{|L|}$$

\tinysection{Generalized \Errorname}
Similarly, the maximum entropy distribution $\overline{\rho}_S$ for the whole log is:
$$\overline{\rho}_S(\vec{q})=\mysuml_{i=1,\ldots,K}w_i* \overline{\rho}_{S_i}(\vec{q})$$
We define the \textit{Generalized \errorname} of a pattern mixture encoding similarly, as the weighted sum of the errors for each partition:
{\small
$$\entropy(\overline{\rho}_S)-\entropy(\rho^*)=\mysuml_{i}w_i \cdot \entropy(\overline{\rho}_{S_i})-\mysuml_{i}w_i\cdot \entropy(\rho^*_i)=\mysuml_{i}w_i \cdot e(S_i)$$
}
As in the base case, a pattern mixture encoding with low Generalized \errorname indicates a high-fidelity representation of the original log.
A process can infer the probability of any query $p(Q=\vec q\;|\;L)$ drawn from the original distribution, simply by inferring its probability drawn from each cluster $i$ (i.e., $p(Q=\vec q\;|\;L_i)$) and taking a weighted average over all inferences. 
When it is clear from context, we refer to Generalized \errorname simply as Error in the rest of this paper.

\tinysection{Generalized Verbosity}
We generalize verbosity to mixture encodings as the \textit{Total Verbosity} ($\sum_{i}|S_i|$), or the total size of the encoded representation.
This approach is ideal for our target applications, where our aim is to reduce the representational size of the query log.

\begin{figure*}[h!]
\captionsetup[subfigure]{justification=centering}
    \centering 
    
\begin{subfigure}[b]{1\textwidth}
    \centering      
    \includegraphics[width=1\textwidth]{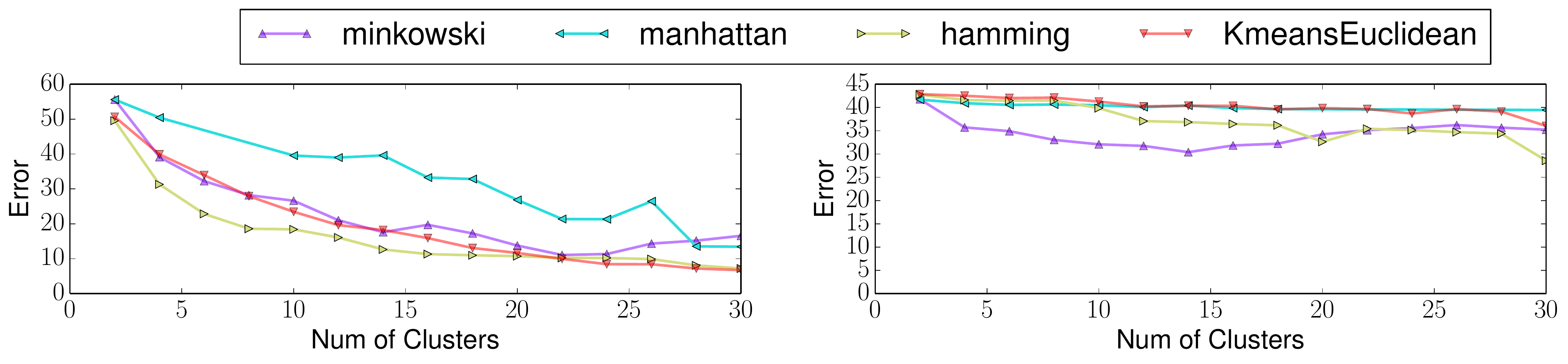}
 \bfcaption{Error v. Number of Clusters (Left: PocketData, Right: US Bank)}   \label{fig:ErrorVNumCluster}
\end{subfigure}
~
\begin{subfigure}[b]{0.986\textwidth}
    \centering      
    \includegraphics[width=1\textwidth]{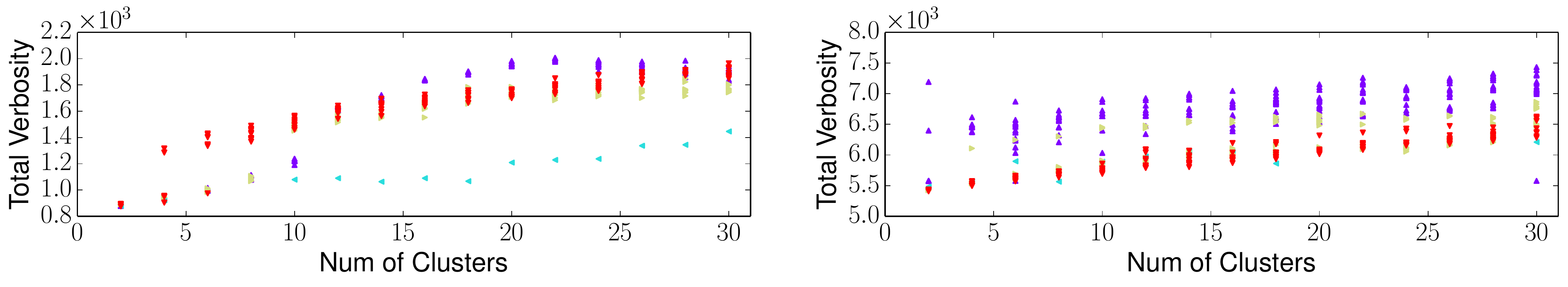}
 \bfcaption{Total Verbosity v. Number of Clusters (Left: PocketData, Right: US Bank); 
Each point is the verbosity of one of ten trials.  The Y-axis' lower bound is the verbosity at 1 cluster to better show the change in verbosity.}   
 \label{fig:TotalVerbosityVNumCluster}
\end{subfigure}
~
\begin{subfigure}[b]{1.001\textwidth}
    \centering      \includegraphics[width=1\textwidth]{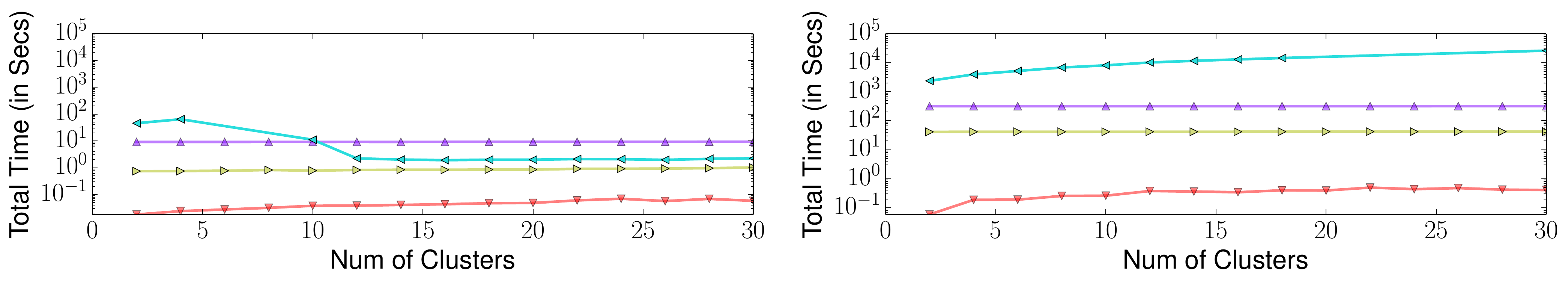}
 \bfcaption{Runtime v. Number of Clusters (Left: PocketData, Right: US Bank)}   \label{fig:RunningTimeVCluster}
\end{subfigure}
~
 \bfcaption{Distance Measure Comparison}
 \label{fig:distancemeasurecomparison} 
 \trimfigurewhitespace
\end{figure*}

\section{Pattern Mixture Compression}
\label{sec:constructingencodings}
We are now ready to describe the \systemname compression scheme.
Broadly, \systemname attempts to identify a pattern mixture encoding that optimizes for some target trade-off between Total Verbosity and Error.  
A naive --- though impractical --- approach to finding such an encoding would be to search the entire space of possible pattern mixture encodings.
Instead, \systemname approximates the same outcome by identifying the \emph{naive} pattern mixture encoding that is closest to optimal for the desired trade-off.
As we show experimentally, the naive mixture encoding produced by the first stage is competitive with more complicated, slower techniques for summarizing query logs.
We also explore a hypothetical second stage, where \systemname refines the naive mixture encoding to further reduce error.
The outcome of this hypothetical stage has a slightly lower Error and Verbosity, but does not admit efficient computation of database statistics. 

\subsection{Constructing Naive Mixture Encodings}
\label{sec:constructingnaivemixtureencodings}
\systemname compression searches for a naive mixture encoding that best optimizes for a requested tradeoff between Total Verbosity and Error.  
As a way to make this search efficient, we observe that a log (or log partition) uniquely determines its naive mixture encoding. 
Thus the problem of searching for a naive mixture encoding reduces to the problem of searching for the corresponding log partitioning.

We further observe that the Error of a naive mixture encoding is proportional to the diversity of the queries in the log being encoded; 
The more uniform the log (or partition), the lower the corresponding error.
Hence, the partitioning problem further reduces to the problem of clustering queries in the log by feature overlap.

To identify a suitable clustering scheme, we next evaluate four commonly used partitioning/clustering methods: (1) KMeans~\cite{Jain:2010:DCY:1755267.1755654} with Euclidean distance (i.e., $l_2$-norm) and Spectral Clustering~\cite{Kannan:2004:CGB:990308.990313} with (2) Manhattan (i.e., $l_1$-norm), (3) Minkowski (i.e., $l_p$-norm) with $p=4$, and (4) Hamming ($\frac{Count(\vec{x}\neq\vec{y})}{Count(\vec{x}\neq\vec{y})+Count(\vec{x}=\vec{y})}$) distances\footnote{We also evaluated Spectral Clustering with Euclidean, Chebyshev and Canberra distances; These did not perform better and we omit them in the interest of conciseness.}.
Specifically, we evaluate these four strategies with respect to their ability to create naive mixture encodings with low Error and low Verbosity.
\tinysection{Experiment Setup}
Spectral and KMeans clustering algorithms are implemented by \textit{sklearn}~\cite{scikit-learn} in Python. 
We gradually increase $K$ (i.e., the number of clusters) configured for each clustering algorithm to mimic the process of continuously sub-clustering the log, tolerating higher Total Verbosity for lower Error.
To compare clustering methods fairly, we reduce randomness in clustering (e.g., random initialization in KMeans) by running each of them $10$ times for each $K$ and averaging the Error of the resulting encodings.  
We used two datasets: ``US Bank'' and ``PocketData.''  
We describe both datsets and the data preparation process in detail in Section~\ref{sec:commonexperimentsettings}. 
All results for our clustering experiments are shown in Figure~\ref{fig:distancemeasurecomparison}.

\subsubsection{Clustering}
\label{sec:clustering}
We next show that clustering is an effective way to consistently reduce Error, although no one clustering method is ideal along all three of Error, Verbosity, and runtime.

\tinysection{More clusters reduces Error}
Figure~\ref{fig:ErrorVNumCluster} compares the relationship between the number of clusters (x-axis) and Error (y-axis), showing the varying rates of convergence to zero Error for each clustering method.
We observe that adding more clusters does consistently reduce Error for both data sets, regardless of clustering method and distance measures.
We note that the US Bank dataset is significantly more diverse than the PocketData dataset, with respect to the total number of features (See Table~\ref{table:datasummary}) and that more than $30$ clusters may be required for reaching near-zero Error.
In general, Hamming distance converges faster than other methods on PocketData.

\tinysection{Adding more clusters increases Verbosity}
Figure~\ref{fig:TotalVerbosityVNumCluster} compares the relationship between the number of clusters (x-axis) and Verbosity (y-axis).
We observe that Verbosity increases with the number of clusters.
This is because when a partition is split, features common to both partitions each increase the Verbosity by 1 each.

\tinysection{Hierarchical Clustering}
Classical clustering methods produce non-monotonic cluster assignments. That is, the ratio of Error to Verbosity can grow with more clusters, as seen in Figure~\ref{fig:ErrorVNumCluster} and~\ref{fig:TotalVerbosityVNumCluster}. 
An alternative is to use hierarchical clustering~\cite{Johnson1967}, which forces monotonic assignments and offers more dynamic control over the Error/Verbosity tradeoff.


\tinysection{Run Time Comparison}
The total run time (y-axis) in Figure~\ref{fig:RunningTimeVCluster} 
includes both distance matrix computation time (if any) and clustering time. 
Note the log-scale: K-Means is orders of magnitude faster than the others, with Hamming distance also performing competitively.

\tinysection{Take-Aways}
Hamming distance provides the best tradeoff between Error and runtime.
For time-sensitive applications, KMeans is preferred to Spectral Clustering.

\tinysection{Visualizing Naive Mixture Encoding}
As with normal pattern summaries, naive mixture summaries are also interpretable.
For example a visualization like that of Figure~\ref{fig:screenshots:nocorrelation} can be repeated, once for each cluster.  
For more details, see Appendix~\ref{appendix:naivemixturesummaryvisualization}.


\subsection{Approximating Log Statistics}
Recall that our primary goal is estimating statistical properties.
In particular, we are interested in counting the occurrences (i.e., the marginal) of some pattern $\pattern$ in the log:
$$\Gamma_\pattern(L) = |\comprehension{\vec q}{\vec q \in L \wedge \pattern \subseteq \vec q}|$$
Recall that a naive encoding $\encoding_L$ includes only single-feature patterns.
Assuming log distributions allowed by this encoding are equally likely, the maximal entropy distribution $\overline \rho_{\encoding\text{\tiny L}}$ 
is the representative distribution.
Hence, we estimate $\Gamma_\pattern(L)$ by multiplying the probability of the feature occurring by the size of the log: $|L| \cdot \overline \rho_{\encoding\text{\tiny L}}$, or:
$$est[\;\Gamma_\pattern(L) \;|\; \encoding_L\;] = |L| \cdot \left(\prod_{f \in \encoding\text{ \textbf{where} }f \subseteq \pattern} \encoding[f]\right)$$

This process trivially generalizes to naive pattern mixture encodings by mixing distributions.  
Specifically, given a set of partitions $L_1 \cup \ldots \cup L_K = L$, the estimated counts for $\Gamma_\pattern(L)$ under each individual partition $L_i$ can be computed based on the partition's encoding $\encoding_i$ and we sum up the estimated counts in each partition
$$est[\;\Gamma_\pattern(L_i)\;|\;\encoding_1, \ldots, \encoding_K\;] = \sum_{i \in [1, K]} est[\;\Gamma_\pattern(L_i)\;|\;\encoding_i\;]$$

\begin{figure*}[h!]
\captionsetup[subfigure]{justification=centering}
    \centering 
    
\begin{subfigure}[b]{0.49\textwidth}
    \centering     
     \includegraphics[width=1\textwidth]{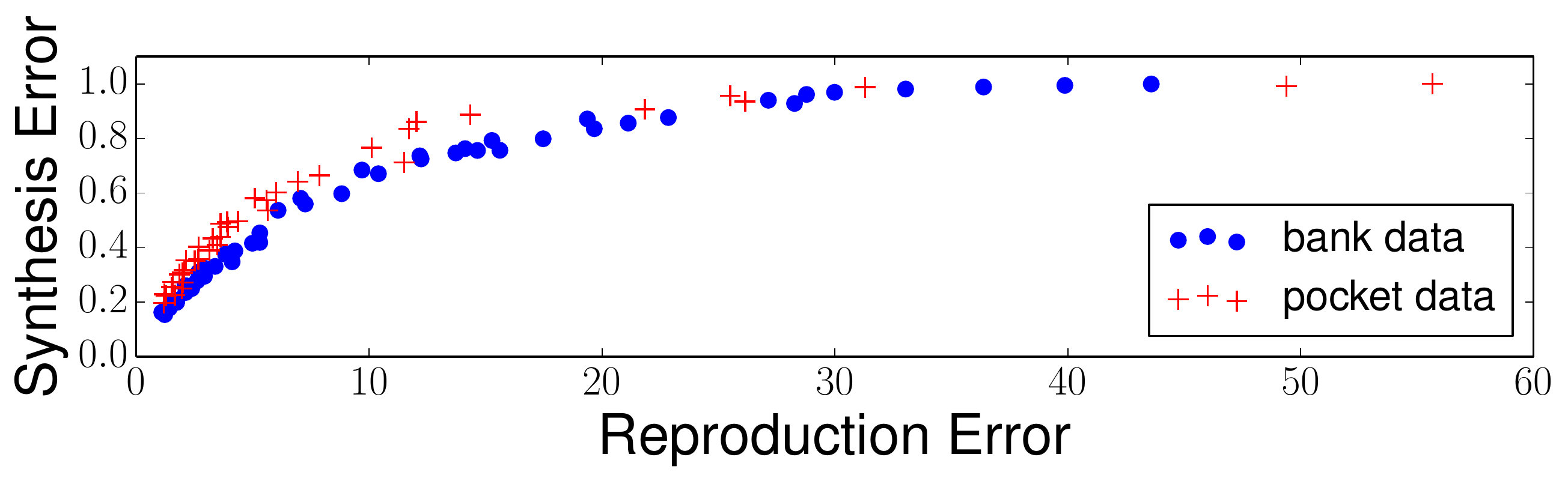}
 \bfcaption{Synthesis Error v. \Errorname}   
 \label{fig:synthesis_error_versus_reproduction_error}
\end{subfigure}
~
\begin{subfigure}[b]{0.49\textwidth}
    \centering      
    \includegraphics[width=1\textwidth]{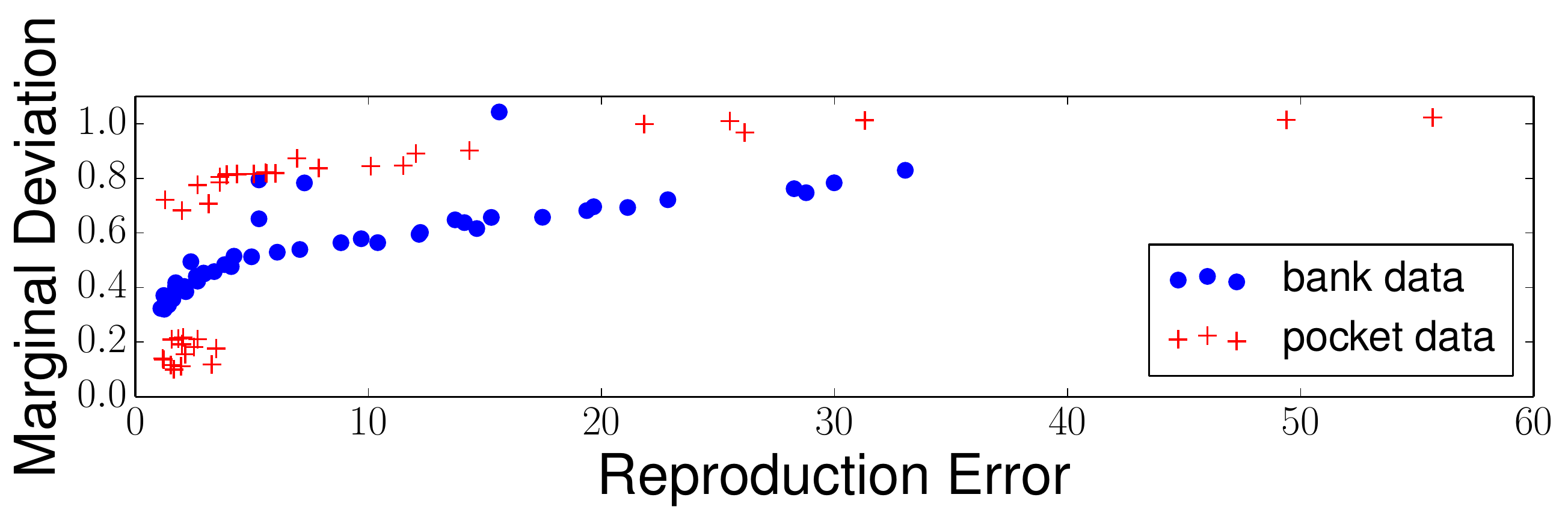}
 \bfcaption{Marginal Deviation v. \Errorname}   
 \label{fig:marginal_deviation_versus_reproduction_error}
\end{subfigure}

 \bfcaption{Effectiveness of Naive Mixture Encoding}
 \label{fig:effectiveness_of_naive_mixture_encoding} 
 \trimfigurewhitespace
\end{figure*} 

\subsection{Pattern Synthesis \& Marginal Estimation}
In this section, we empirically verify the effectiveness of naive mixture encodings in approximating log statistics from two related perspectives. 
The first perspective focuses on \textit{synthesis error}. 
It measures whether patterns synthesized by the naive mixture encoding actually exist in the log.
From the second perspective, we would like to further investigate the \textit{marginal deviation} of patterns contained in the log of queries.
This evaluates whether a naive mixture encoding will compute the correct marginal for patterns of interest to a client application.
Specifically, \textit{synthesis error} is measured as $1-\frac{M}{N}$ where $N$ is the total number of randomly synthesized patterns and $M$ is the number of synthesized patterns with positive marginals in the log.
\textit{Marginal deviation} is measured as $\frac{|ESTM-TM|}{TM}$ where $TM$ stands for True Marginal of a pattern and $ESTM$ is the one estimated by naive mixture encoding.

Experimental results are shown in Figure~\ref{fig:effectiveness_of_naive_mixture_encoding}.
Both synthesis error and marginal deviation consistently decreases given more clusters.
Furthermore, as we vary the number of clusters, both measures are correlated with \errorname.

\tinysection{Synthesis Error}
Figure~\ref{fig:synthesis_error_versus_reproduction_error} shows synthesis error (y-axis) versus \errorname (x-axis).
The figure is generated by randomly synthesizing $N=10000$ patterns from each partition of the log.
Note that different values of $N$ give similar observation.
The overall synthesis error is measured as the average synthesis error for each partition, weighted by proportion of queries in the partition.

\tinysection{Marginal Deviation}
Figure~\ref{fig:marginal_deviation_versus_reproduction_error} shows marginal deviation (y-axis) versus \errorname (x-axis).
It is not feasible to enumerate all patterns that exist in the data.
As an alternative, we treat each distinct query in the log as a pattern and treat the marginal deviation on it as the worst case for all patterns that it may contain. This is because marginal deviation tends to be smaller if it is measured on a pattern that is contained in the other.
For each cluster, we sum up the marginal deviation on all distinct queries and the final marginal deviation for the whole log is an weighted average (same as synthesis error) over all clusters.

\subsection{Naive Encoding Refinement}
\label{sec:naivemixtureencodingrefinement}
Naive mixture encodings can already achieve close to near-zero Error (Figure~\ref{fig:ErrorVNumCluster}), have low Verbosity, and admit efficiently computable log statistics $\Gamma_\pattern(L)$.
Although doing so makes estimating statistics more computationally expensive, as a thought experiment, we next consider how much of an improvement we could achieve in the Error/Verbosity tradeoff by exploring a hypothetical second stage that enriches naive mixture encodings by adding non-naive patterns.

\tinysection{Feature-Correlation Refinement}
The first challenge is that our closed-form formula for \errorname only works for naive encodings.
Hence, we first consider the simpler problem of identifying the individual pattern that most reduces the \errorname of a naive encoding.

Recall that the closed-form representation for the \errorname arises by independence between features (i.e., $\overline \rho_S(Q = \vec q) = \prod_{i} p(X_i = x_i)$).
Similarly, under naive encodings we have a closed-form estimation of marginals $p(Q\supseteq\vec b)$ (i.e., $\overline \rho_S(Q \supseteq \vec b) = \prod_{i} p(X_i \geq x_i)$).
We define the \textit{feature-correlation} of pattern $\vec b$ as the log-difference from its actual marginal to the estimation, according to naive encoding.
$$WC(\vec b, S) = \log\left(
    p(Q \supseteq \vec b)
\right) - \log\left(
  \overline \rho_S(Q \supseteq \vec b)
\right)$$
Intuitively, patterns with higher feature correlations create higher Errors, which in turn makes them ideal candidates for addition to the compressed log encoding.
For two patterns with the same feature-correlation, the one that occurs more frequently will have greater impact on Error~\cite{Han:2007:FPM:1275092.1275097}. 
As a result, we compute an overall score for ranking patterns involving feature-correlation:  
$$corr\_rank(\vec{b})=p(Q \supseteq \vec b) \cdot WC(\vec b, S)$$
We show in Section~\ref{sec:motivateencodingerror} that $corr\_rank$ closely correlates with \errorname.
That is, a higher $corr\_rank$ value indicates that a pattern produces a greater \errorname reduction if introduced into the naive encoding.

\tinysection{Pattern Diversification}
This greedy approach only allows us to add a single pattern to each cluster.
In general, we would like to identify a \textit{set} of patterns.
We cannot sum up the $corr\_rank$ of each pattern in the set to estimate its \errorname, as information content carried by patterns may overlap.
To counter such overlap, or equivalently to \textit{diversify} patterns, a search through the space of pattern sets is needed.
This type of diversification is commonly used in pattern mining applications, but can quickly become expensive.
As we show experimentally in Section~\ref{sec:motivatepatternmixturesummaries}, the benefit that can be obtained from diversification is minimal.

\section{Experiments}
\label{sec:experiments}
In this section, we design experiments to empirically (1)~validate that \errorname correlates with Deviation and (2)~evaluate the effectiveness of \systemname compression.

We use two specific datasets in the experiment: (1) SQL query logs of the Google+ Android app extracted from the PocketData public dataset~\cite{kennedy:2015:tpc-tc:pocket} and (2) SQL query logs that capture all query activity on the majority of databases at a major US bank over a period of approximately 19 hours.
A summary of these two datasets is given in Table~\ref{table:datasummary}.

\begin{table}
\centering
\bfcaption{Summary of Data sets}
\label{table:datasummary}
{\small \centering
\begin{tabular}{c c c}
\toprule
Statistics & PocketData & US bank \\
\midrule
\# Queries & 629582& 1244243\\
\midrule
\# Distinct queries & 605& 188184\\
\midrule
\# Distinct queries (w/o const)& 605& 1712\\
\midrule
\# Distinct conjunctive queries & 135& 1494\\
\midrule
\# Distinct re-writable queries & 605& 1712\\
\midrule
Max query multiplicity & 48651 & 208742\\
\midrule
\# Distinct features & 863& 144708\\
\midrule
\# Distinct features (w/o const) & 863& 5290\\
\midrule
Average features per query & 14.78& 16.56\\
\bottomrule
\end{tabular}
}
\trimfigurewhitespace
\end{table}

\begin{figure*}[h!]
	\captionsetup[subfigure]{justification=centering}
    \centering
    \begin{subfigure}[b]{0.45\textwidth}
        \centering
        \includegraphics[width=\textwidth]{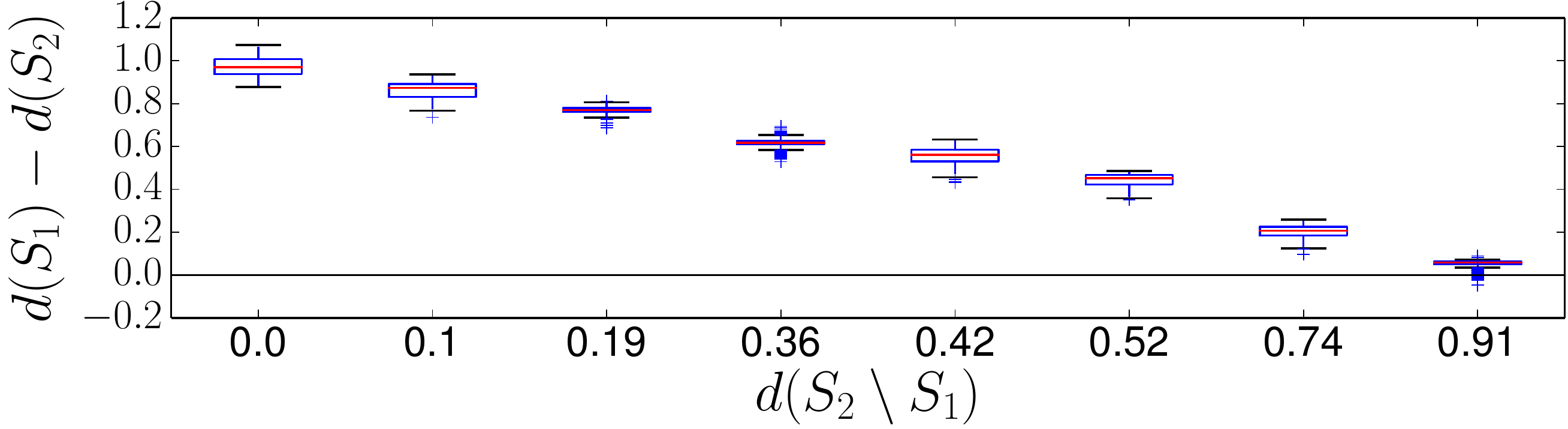}
        \bfcaption{Containment captures Deviation (US bank)}
        \label{fig:containmentcapturesdeviation_bankdata}
    \end{subfigure}
        ~
    \begin{subfigure}[b]{0.45\textwidth}
        \centering
        \includegraphics[width=\textwidth]{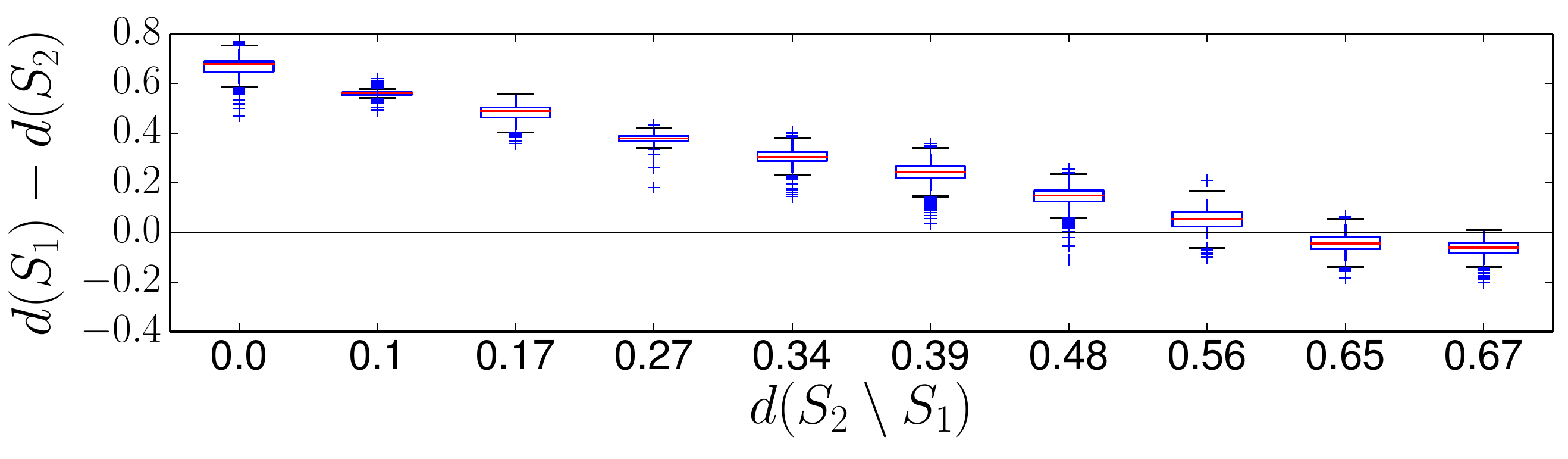}
        \bfcaption{Containment captures Deviation (PocketData)}
        \label{fig:containmentcapturesdeviation_pocketdata}
    \end{subfigure}
    ~
    \begin{subfigure}[b]{0.45\textwidth}
        \centering
         \includegraphics[width=\textwidth]{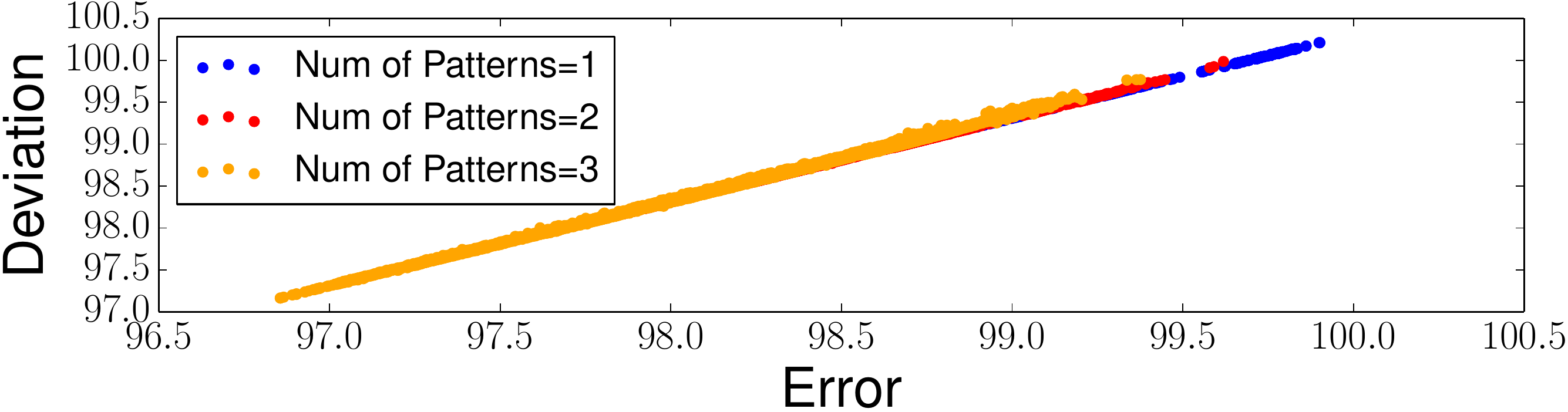}
        \bfcaption{Error captures Deviation (US bank)}
        \label{fig:errorcapturesdeviation_bankdata}
    \end{subfigure}
        ~
    \begin{subfigure}[b]{0.45\textwidth}
        \centering
        \includegraphics[width=\textwidth]{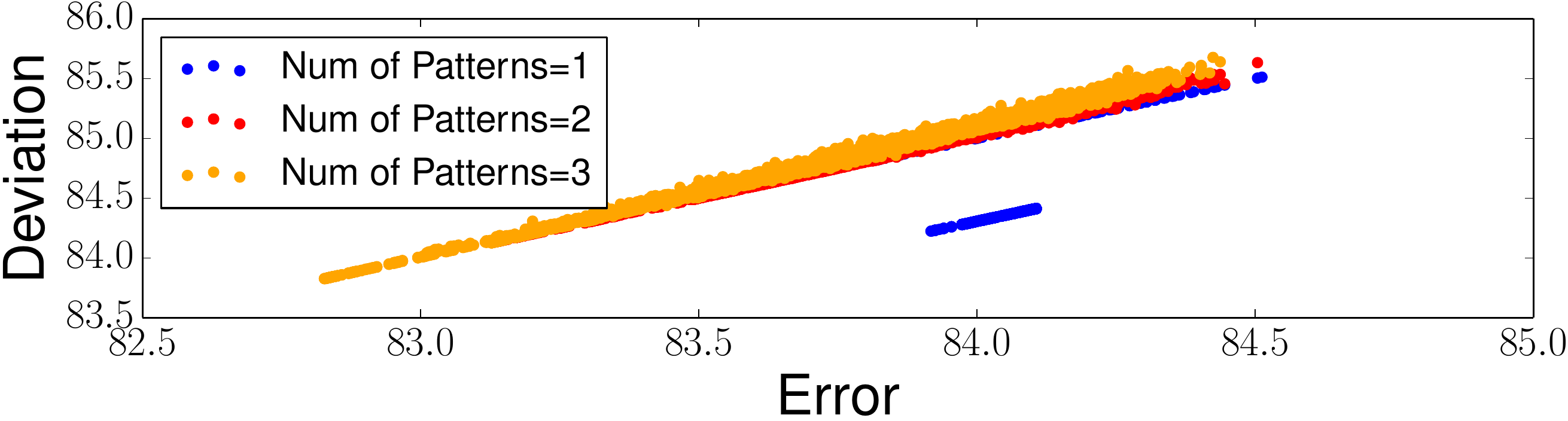}
        \bfcaption{Error captures Deviation (PocketData)}
        \label{fig:errorcapturesdeviation_pocketdata}
    \end{subfigure}
    ~
     \begin{subfigure}[b]{0.45\textwidth}
        \centering     \includegraphics[width=\textwidth]{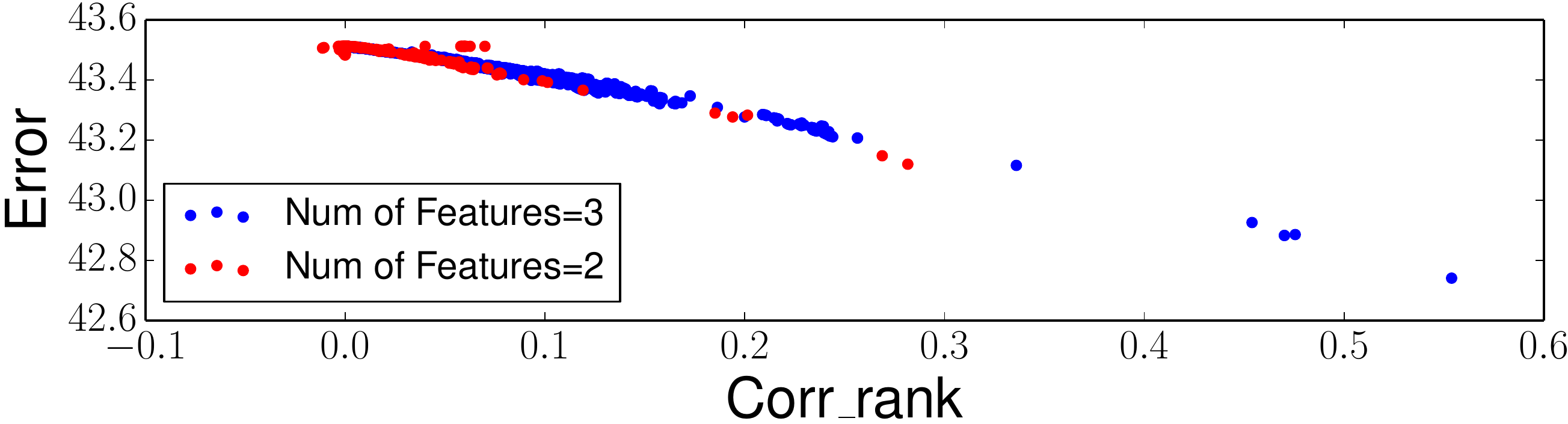}
        \bfcaption{Error captures Correlation (US bank)}
        \label{fig:errorcapturescorrelation_bankdata}
\end{subfigure}
    ~
     \begin{subfigure}[b]{0.45\textwidth}
        \centering     \includegraphics[width=\textwidth]{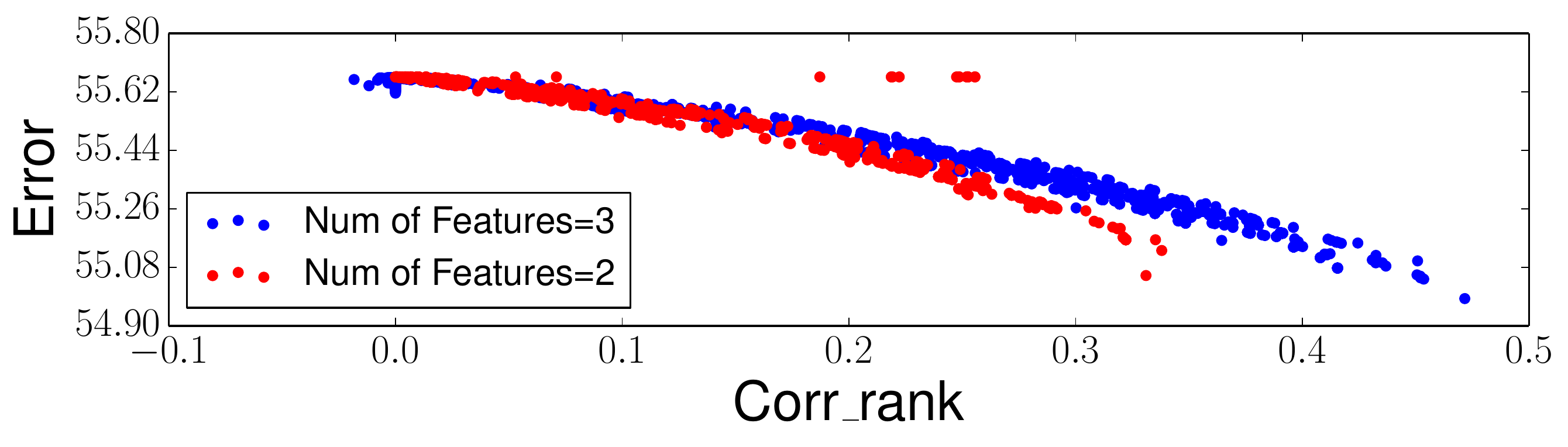}
        \bfcaption{Error captures Correlation (PocketData)}
        \label{fig:errorcapturescorrelation_pocketdata}
\end{subfigure}
    
\bfcaption{Validating The \Errorname Metric}
\label{fig:validatingencodingerror}
\trimfigurewhitespace
\end{figure*}

\tinysection{The PocketData-Google+ query log} The dataset consists of SQL logs that capture all database activities of 11 Android phones. 
We selected Google+ application for our study since it is one of the few applications where all users created a workload. 
This dataset can be characterized as a stable workload of exclusively machine-generated queries.

\tinysection{The US bank query log}
This log is an anonymized record of queries processed by multiple relational database servers at a major US bank~\cite{Kul:2016:EAQ:2872518.2888608} over a period of 19 hours.
Of the nearly 73 million database operations captured, 58 million are not directly queries, but rather invocations of stored procedures and 13 million not able to be parsed by standard SQL parser. 
Among the rest of the 2.3 million parsed SQL queries, since we are focusing on conjunctive queries, we base our analysis on the 1.25 million valid \texttt{SELECT} queries. 
This dataset can be characterized as a diverse workload of both machine- and human-generated queries.

\tinysection{Common Experiment Settings}
\label{sec:commonexperimentsettings}
Experiments were performed on a 2.8 GHz Intel Core i7 CPU with 16 GB 1600 MHz DDR3 memory and a SSD running macOS Sierra.

\tinysection{Constant Removal}
A number of queries in US Bank differ only in hard-coded constant values.
Table~\ref{table:datasummary} shows the total number of queries, as well as the number of distinct queries if we ignore constants.
By comparison, queries in PocketData all use JDBC parameters.
For these experiments, we ignore constant values in queries.

\tinysection{Query Regularization}
We apply query rewrite rules (similar to \cite{Chandra:2016:PMA:3007263.3007304}) to regularize queries into equivalent conjunctive forms, where possible. 
Table~\ref{table:datasummary} shows that $\frac{135}{605}$ and $\frac{1494}{1712}$ of distinct queries are in conjunctive form for PocketData and US bank respectively. 
After regularization, all queries in both data sets can be either simplified into conjunctive queries or re-written into a \texttt{UNION} of conjunctive queries compatible with Aligon et. al.'s feature scheme~\cite{Aligon2014}.

\tinysection{Convex Optimization Solving}
All convex optimization problems involved in measuring \errorname and Deviation are solved by the \textit{successive approximation heuristic} implemented by the CVX toolbox~\cite{cvx} with Sedumi solver.

\subsection{Validating \Errorname}
\label{sec:motivateencodingerror}
In this section, we validate that \errorname is a practical alternative to Deviation.
In addition, we also offer measurements on its correlation with Deviation, as well as feature-correlation described in Section~\ref{sec:naivemixtureencodingrefinement}.

Since Deviation cannot be measured exactly (See Section~\ref{sec:idealizedrepresentativenessmeasures}), we approximate it using sampling, which is further explained in Appendix~\ref{appendix:sampling}.
It is impractical to enumerate all possible encodings, we choose a subset of encodings for both datasets. 
Specifically, we first select all features with marginals in the range $[0.01,0.99]$ and use these features to construct patterns.
We then enumerate combinations of $K$ (up to 3) patterns as our chosen encodings.

\tinysection{Containment Captures Deviation}
Here we empirically verify that containment (Section~\ref{sec:validateencodingerror}) captures Deviation (i.e., $\encoding_1\leq_{\Omega} \encoding_2\to d(\encoding_1)\leq d(\encoding_2)$) to complete the chain of reasoning that \errorname captures Deviation.
Figures~\ref{fig:containmentcapturesdeviation_bankdata} and \ref{fig:containmentcapturesdeviation_pocketdata} show all pairs of encodings where $\encoding_2\supset \encoding_1$.
The y-axis shows the difference in Deviation values (i.e., $d(\encoding_2)-d(\encoding_1)$).
Deviation $d(S)$ is approximated by drawing 1,000,000 samples from the space of possible patterns.
For clarity, we bin pairs of encodings by the degree of overlap between the encodings, measured by the Deviation of the set-difference between the two encodings $d(\encoding_2\setminus \encoding_1)$; Higher $d(\encoding_2\setminus \encoding_1)$ implies less overlap. 
Y-axis values are grouped into bins and visualized by boxplot (i.e., the blue box indicates the range within standard deviation and red/black crosses are outliers).
Intuitively, all points above zero on the y-axis (i.e., $d(\encoding_2)-d(\encoding_1) > 0$) are pairs of encodings where Deviation order agrees with containment order.
This is the case for virtually all encoding pairs.  

\tinysection{Additive Separability of Deviation}
We also observe from Figures~\ref{fig:containmentcapturesdeviation_bankdata} and ~\ref{fig:containmentcapturesdeviation_pocketdata} that agreement between Deviation and containment order is correlated with overlap; More similar encodings are more likely to have agreement.
Combined with Proposition~\ref{prop:monotone}, this shows first that for similar encodings, \errorname is likely to be a reliable indicator of Deviation.
This also suggests that Deviation is additively separable: The information loss (measured in $d(\encoding_2)-d(\encoding_1)$) by excluding encoding $\encoding_2\setminus \encoding_1$ from $\encoding_2$ closely correlates with the quality (i.e., $d(\encoding_2\setminus \encoding_1)$) of encoding $\encoding_2\setminus \encoding_1$ itself:\vspace*{-4mm}

{\small
$$\encoding_2\supset \encoding_1\to d(\encoding_2)-d(\encoding_1)<0\hspace{5mm}\text{and}\hspace{5mm}d(\encoding_2\setminus \encoding_1)\approx d(\encoding_2)-d(\encoding_1)$$ 
}\vspace*{-5mm}

\tinysection{Error correlates with Deviation}
As a supplement, Figures~\ref{fig:errorcapturesdeviation_bankdata} and ~\ref{fig:errorcapturesdeviation_pocketdata} empirically confirm that that \errorname (x-axis) indeed closely correlates with Deviation (y-axis).
Mirroring our findings above, correlation between them is tighter at lower \errorname.


\tinysection{Error and Feature-Correlation}
Figure~\ref{fig:errorcapturescorrelation_bankdata} and ~\ref{fig:errorcapturescorrelation_pocketdata} show the relationship between \errorname (y-axis) and the feature-correlation score $corr\_rank$ (x-axis), as defined in Section~\ref{sec:naivemixtureencodingrefinement}. Values of y-axis are computed from the naive encoding extended by a single pattern $\vec{b}$ containing multiples features (up to 3). 
One can observe that the \errorname of extended naive encodings almost linearly correlates with $corr\_rank(\vec b)$. 
In addition, one can also observe that $corr\_rank$ becomes higher when the pattern $\vec{b}$ encodes more correlated features. 

\subsection{Feature-Correlation Refinement}
\label{sec:motivatepatternmixturesummaries}
In this section, we design experiments serving two purposes: (1) Evaluating the potential reduction in Error from refining naive mixture encodings through state-of-the-art pattern based summarizers, and (2) Evaluating whether we can replace naive mixture encodings by the encodings created from summarizers that we have plugged-in.

\tinysection{Experiment Setup}
To serve both purposes, we construct pattern mixture encodings under three different configurations: (1) Naive mixture encoding; (2) Pattern based encoding and (3) Naive mixture encoding refined by pattern based encoding.
Naive mixture encodings are constructed by KMeans clustering. 
Pattern based encodings are generated by two state-of-the-art pattern based summarizers: 
(1) \textit{Laserlight}~\cite{ElGebaly:2014:IIE:2735461.2735467} algorithm, which aims at summarizing multi-dimensional data $D=(X_1,\ldots,X_n)$ augmented with an additional binary attribute $A$; 
(2) \textit{MTV}~\cite{Mampaey:2012:SDS:2382577.2382580} algorithm, which aims at mining maximally informative patterns that summarize multi-dimensional binary data. 

The experiment results are shown in Figure~\ref{fig:motivatenaivemixtureencodings_bankdata} which contains 3 sub-figures.
All sub-figures share the same x-axis, i.e., the number of clusters.
Figure~\ref{fig:PatternMixtureEncodingErrorComparisonPiggybacking_bankdata} evaluates the possible change in Error (y-axis) by plugging-in \textit{MTV} and \textit{Laserlight}.
Figure~\ref{fig:PatternMixtureEncodingErrorComparisonAlone_bankdata} compares the Error (y-axis) between the naive mixture encoding and the pattern mixture encoding obtained from only using patterns from \textit{MTV} and \textit{Laserlight}.
Figure~\ref{fig:mixtureencodingsrunningtimecomparison_bankdata} compares the running time (y-axis) between constructing naive mixture encodings and applying pattern based summarizers.
We only show the results for US bank data set as results for PocketData give similar observations. 

\subsubsection{Pattern vs Naive Pattern Mixture Encodings}
\label{sec:Replacing_Naive_Mixture_Encodings}
\begin{figure}[ht!]
	\captionsetup[subfigure]{justification=centering}
    \centering
    \begin{subfigure}[b]{0.48\textwidth}
        \centering       
        \includegraphics[width=\textwidth]{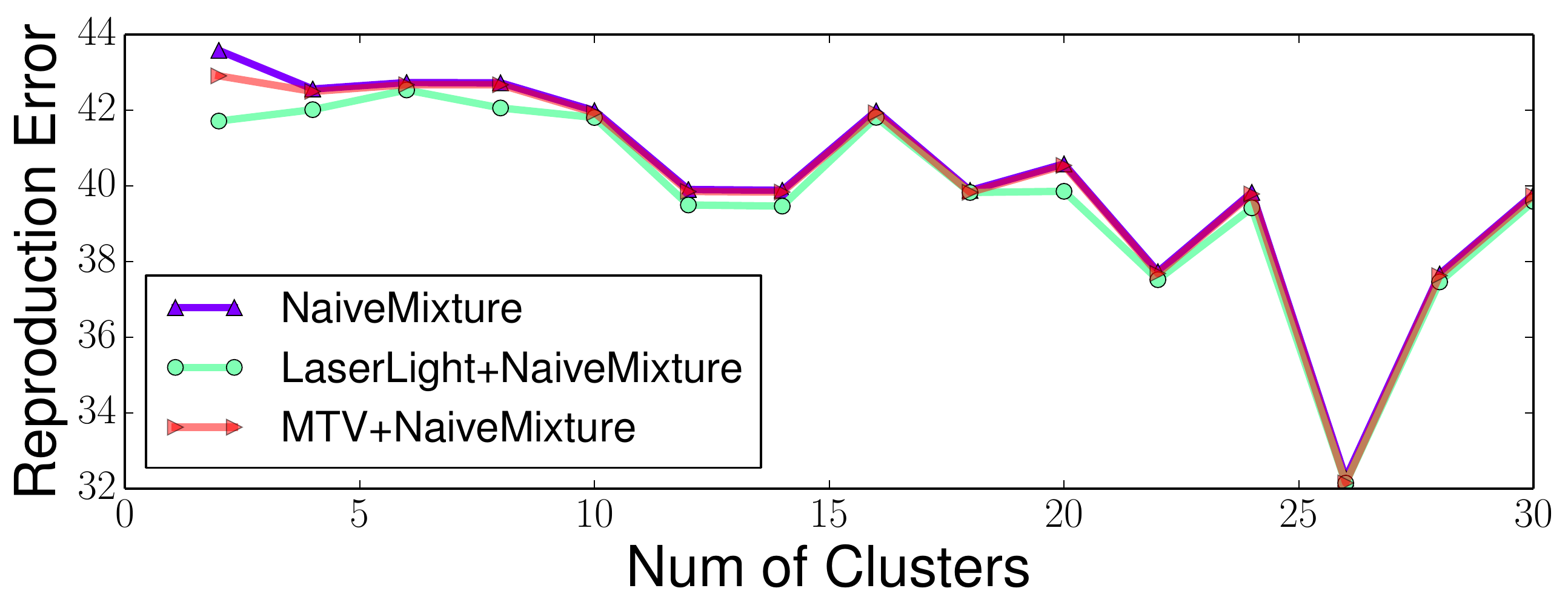}
 \bfcaption{Naive Mixture v. Naive Mixture+LaserLight/MTV. Note that we y-axis is offset (non-zero start).}      \label{fig:PatternMixtureEncodingErrorComparisonPiggybacking_bankdata}
\end{subfigure}
    ~
\begin{subfigure}[b]{0.48\textwidth}
  \centering       
  \includegraphics[width=\textwidth]{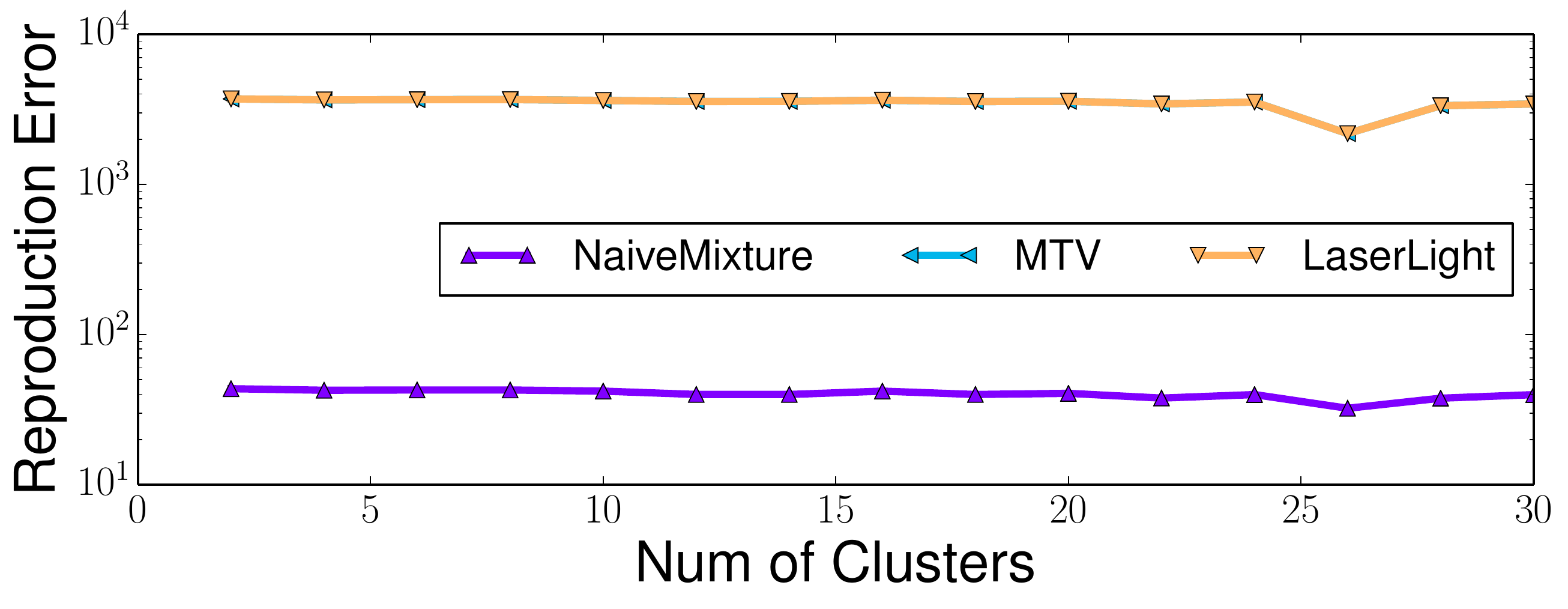}
 \bfcaption{Naive Mixture v. LaserLight/MTV alone. Note that y-axis is in log scale.}     \label{fig:PatternMixtureEncodingErrorComparisonAlone_bankdata}
\end{subfigure}
~
\begin{subfigure}[b]{0.48\textwidth}
\centering       
\includegraphics[width=\textwidth]{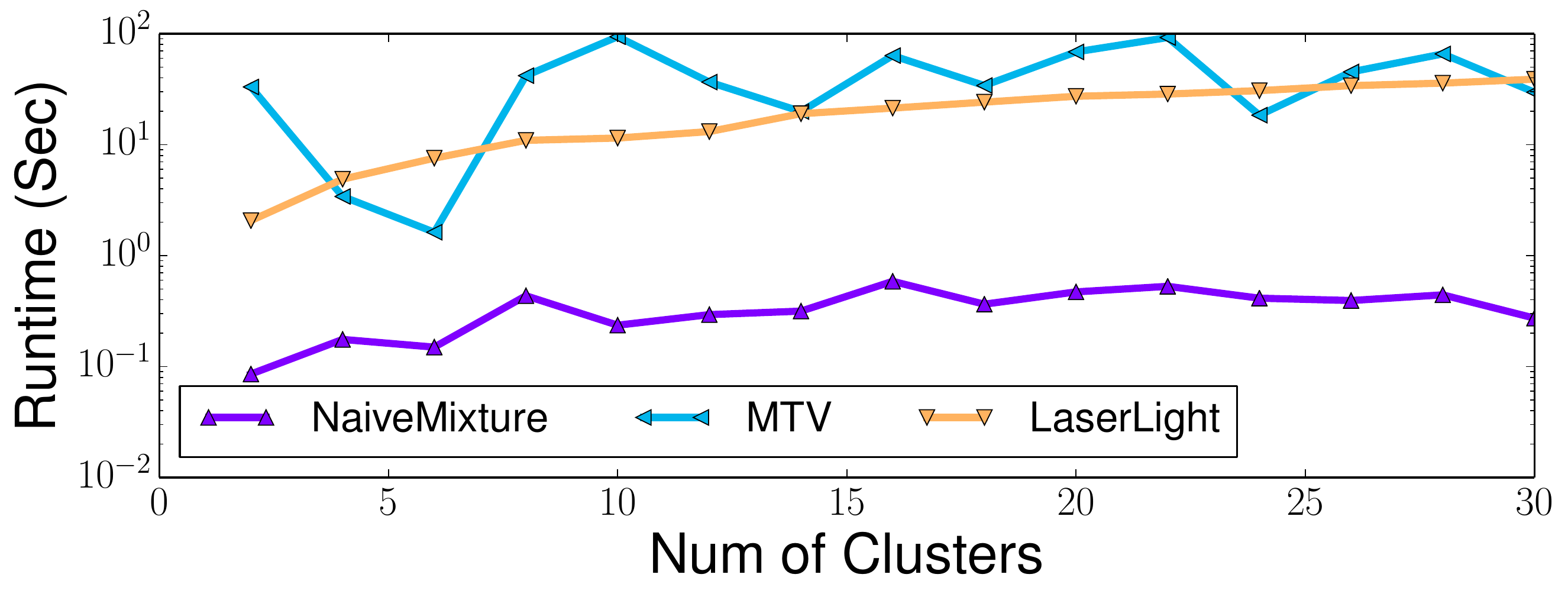}
\bfcaption{Runtime Comparison (y-axis in log scale)}   
\label{fig:mixtureencodingsrunningtimecomparison_bankdata}
\end{subfigure}
\bfcaption{Naive Mixture Encodings (US bank)}   \label{fig:motivatenaivemixtureencodings_bankdata}
\trimfigurewhitespace
\end{figure}

Figure~\ref{fig:PatternMixtureEncodingErrorComparisonAlone_bankdata} and~\ref{fig:mixtureencodingsrunningtimecomparison_bankdata} suggest that naive mixture encodings outperform pattern based encodings in two ways.

\tinysection{Computation Efficiency}
Furthermore, as one can observe from Figure~\ref{fig:mixtureencodingsrunningtimecomparison_bankdata}, that the running time of constructing naive mixture encodings is significantly lower than that of \textit{Laserlight} and \textit{MTV}.

\tinysection{\errorname}
We observe from Figure~\ref{fig:PatternMixtureEncodingErrorComparisonAlone_bankdata} that the \errorname of naive mixture encodings are orders of magnitude lower than those obtained from summarizing using the patterns generated by \textit{Laserlight} or \textit{MTV}.

\tinysection{Verbosity} 
The one way in which pattern based encodings outperform naive pattern mixtures is in verbosity. 
Both \textit{Laserlight} and \textit{MTV} produce encodings with significantly fewer patterns, as the naive pattern mixture summary requires at least one pattern for each feature (e.g., 5290 patterns in the US bank dataset).
Conversely, mining this number of patterns is not computationally feasible (Figure~\ref{fig:mixtureencodingsrunningtimecomparison_bankdata}).


\subsubsection{Refining Naive Mixture Encodings}
\label{sec:refiningnaivemixtureencodings}
The experiment result is shown in Figure~\ref{fig:PatternMixtureEncodingErrorComparisonPiggybacking_bankdata}.
Note that we offset y-axis to show the change in Error.
We observe from the figure that reduction on Error contributed by plugging-in pattern based summarizers is small for both algorithms.

\tinysection{Dimensionality Restriction}
For \textit{Laserlight}, this observation is partially due to the fact that we only keep top $100$ features (in terms of variability) of the data as its input, since \textit{Laserlight} is implemented in PostgresSQL 9.1 which has a threshold of $100$ arguments (one argument for each feature) that can be passed to a function.

\tinysection{Pattern Restriction}
For \textit{MTV}, this is due to a limitation of $15$ patterns that we have experienced in configuring it.
We refer the reader to Section 4.5 of the paper~\cite{Mampaey:2012:SDS:2382577.2382580} that explains the difficulty in inferring the maximum entropy distribution with increasing number of patterns.

\section{Alternative Applications}
\label{sec:evaluatingalternativeapplications}
To fairly evaluate \textit{Laserlight} and \textit{MTV}, we incorporate their own data sets and empirically evaluate them against \textit{naive mixture encoding} under their own applications.

\tinysection{Data Sets}
Specifically, we choose \textit{Mushroom} data set used in \textit{MTV}~\cite{Mampaey:2012:SDS:2382577.2382580} which is obtained from FIMI dataset repository and U.S. Census data on Income or simply \textit{Income} data set, which is downloaded from IPUMS-USA at \textit{https://usa.ipums.org/usa/} and used in \textit{Laserlight}~\cite{ElGebaly:2014:IIE:2735461.2735467}.
The basic statistics of the data sets are given in Table~\ref{table:extendeddatasummary}.

\begin{table}[h!]
\centering
\bfcaption{Data Sets of Alternative Applications}
\label{table:extendeddatasummary}
{\small \centering
\begin{tabular}{c c c}
\toprule
Statistics & Income & Mushroom \\
\midrule
\# Distinct data tuples & 777493 & 8124\\
\midrule
\# Features per tuple & 9 & 21\\
\midrule
Feature Binary-valued? & no& no\\
\midrule
\# Distinct features & 783 & 95\\
\midrule
Binary Classification Feature & $>100,000$? & Edibility\\
\midrule
Assumed data tuple multiplicity & 1 & 1\\
\bottomrule
\end{tabular}
}
\end{table}

\subsection{Experiments}
\label{sec:evaluatingalternativeapplicationsexperiments}
All experiments involving \textit{Laserlight} and \textit{MTV} will be evaluated under their own Error measures and data sets, unless otherwise stated.
The experiments are organized as follows: First, we establish baselines by evaluating classical \textit{Laserlight} and \textit{MTV} on their original data; Then we show that classical \textit{Laserlight} and \textit{MTV} can be generalized to partitioned data and that the generalization improves on their Error measures and also runtime; At last, we compare their generalized versions with \textit{naive mixture encoding} to show that \textit{naive mixture encoding} is a reasonable alternative.

\subsubsection{Error Measures}
We first explain how \textit{naive mixture encoding} is evaluated based on Error defined by \textit{Laserlight} and \textit{MTV}.

\tinysection{Evaluating Naive Encoding on Laserlight Error}
Algorithm \textit{Laserlight} summarizes data $D$ which consists of feature vectors $t$ augmented by some binary feature $v$.
Denote the valuation of the binary feature $v$ for each feature vector $t$ as $v(t)$. 
The goal is to mine a summary encoding $\encoding$, which is a set of patterns contained in $t\in D$ that offer predictive power on $v(t)$.
Denote the estimation (based on $\encoding$) of $v(t)$ as $u_{\encoding}(t)\in [0,1]$, the \textit{Laserlight} Error is measured by $$\sum_t ( v(t)\log(\frac{v(t)}{u_{\encoding}(t)})+(1-v(t))\log(\frac{1-v(t)}{1-u_{\encoding}(t)}) )$$
Since \textit{naive encoding} $\encoding_L$ is equivalent to independence assumption on features, $u_{\encoding_L}(t)$ is simply the probability of $v(t)=1$, namely $u_{\encoding_L}(t)=\frac{|\{t|v(t)=1,t\in D\}|}{|D|}$ regardless of $t$.
Consequently, the \textit{Laserlight} Error of \textit{naive encoding} is $$-|D|(u_{\encoding_L}\log u_{\encoding_L}+(1-u_{\encoding_L})\log (1-u_{\encoding_L}) )$$

\tinysection{Evaluating Naive Encoding on MTV Error}
We denote the data also as $D$ with some summary encoding $\encoding$, the \textit{MTV} Error of $\encoding$ is $$-|D|H(\rho^*)+1/2|\encoding|\log|D|$$ where $H(\rho^*)$ stands for the entropy of maximum entropy distribution $\rho^*$ defined in the paper.
The second term in \textit{MTV} Error penalizes the verbosity of the encoding $\encoding$.
We define the entropy of some feature $f$ as $H(f)=-p\log p-(1-p)\log(1-p)$ where $p$ is the probability of the feature being present, $H(\rho^*)$ of \textit{naive encoding} is simply the summation $\sum_f H(f)$ over all feature entropies.

\tinysection{Evaluating Naive Mixture Encoding}
Evaluation of \textit{naive encoding} can be generalized to \textit{naive mixture} by taking a weighted average over resulting clusters (See Section~\ref{sec:generalizedinformationlossmeasures}).

\begin{figure}[h!]
    \captionsetup[subfigure]{justification=centering}
    \centering
    \begin{subfigure}[b]{0.48\textwidth}
        \centering       
        \includegraphics[width=\textwidth]{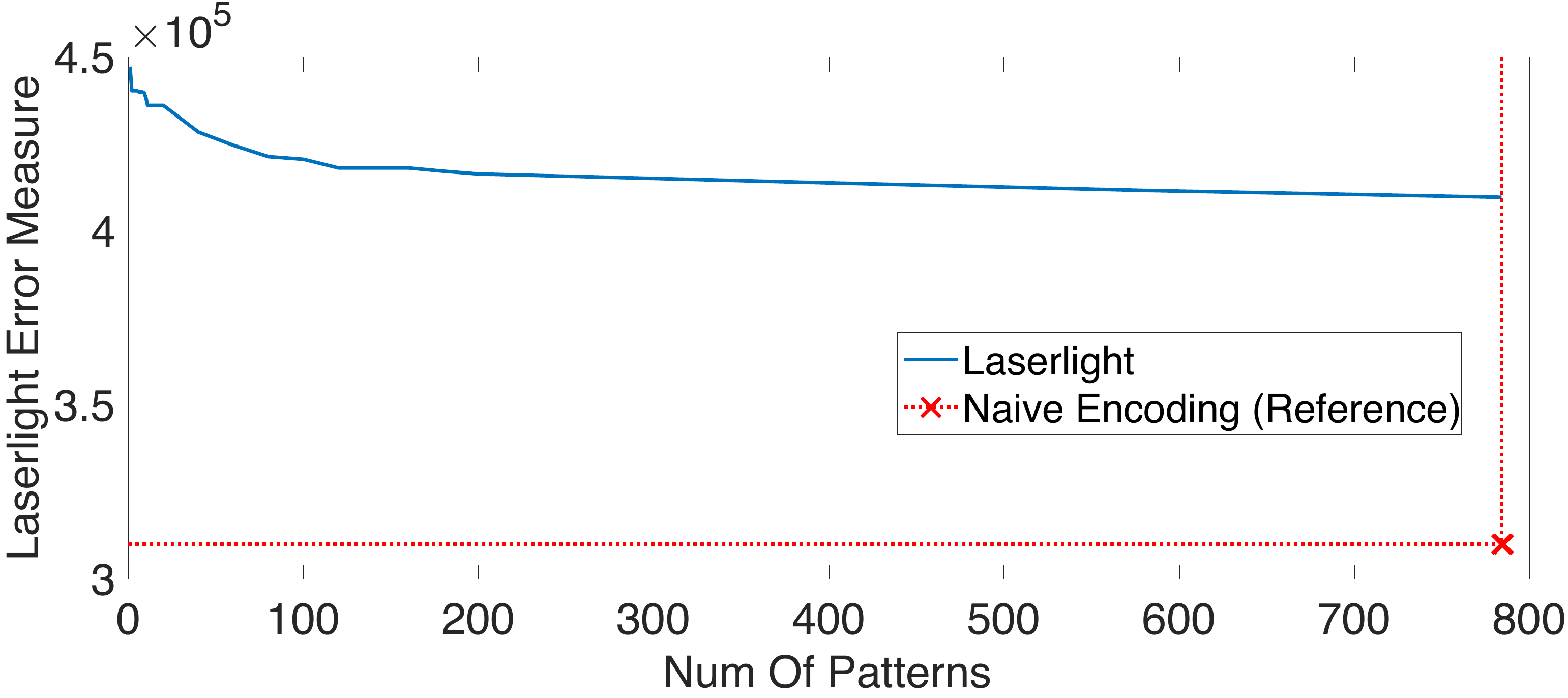}
        \bfcaption{Laserlight Error v. \# of Patterns on Income data}
        \label{fig:Laserlight_Error_vs_NumOfPatterns}
    \end{subfigure}
    ~
    \begin{subfigure}[b]{0.48\textwidth}
        \centering       
        \includegraphics[width=\textwidth]{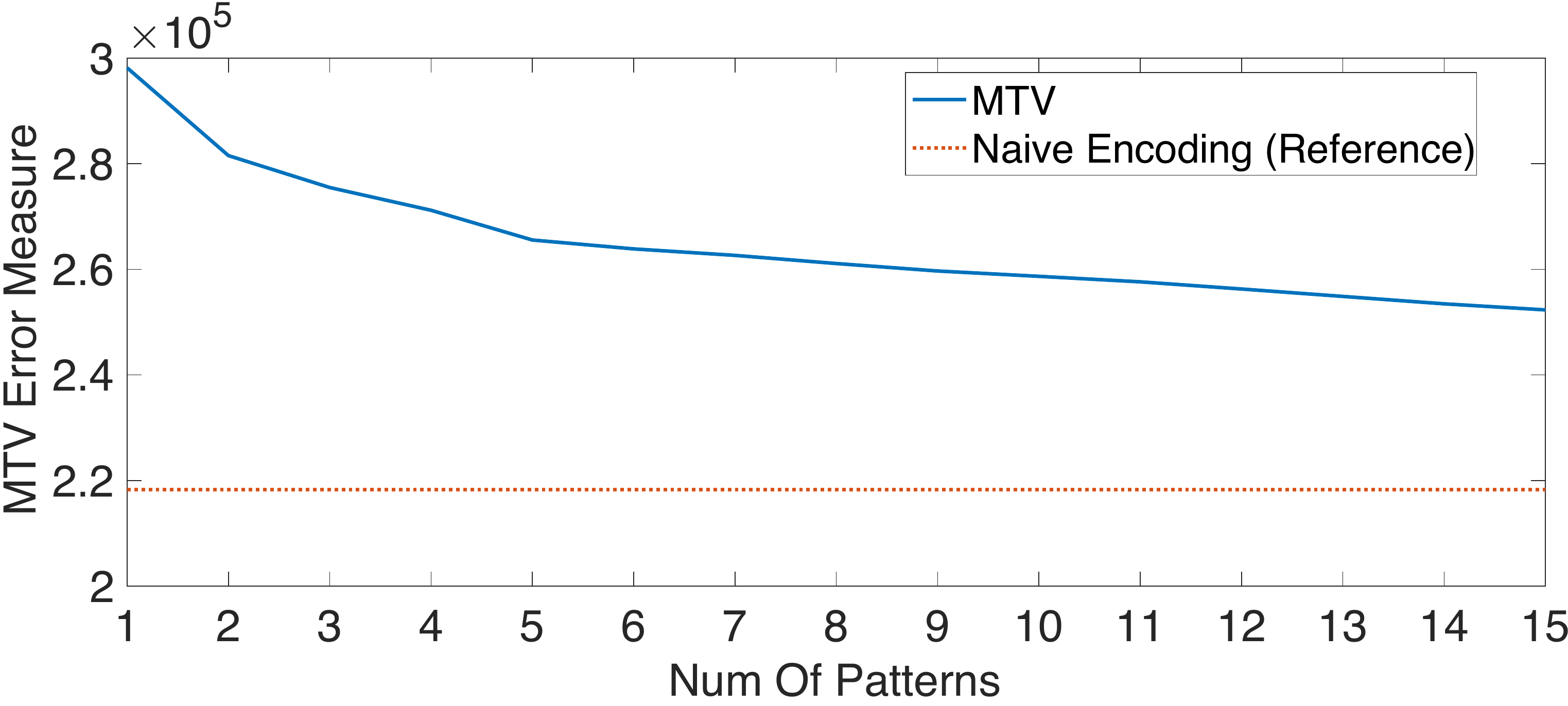}
        \bfcaption{MTV Error v. \# of Patterns on Mushroom data}     
        \label{fig:MTV_Error_vs_NumOfPatterns}
    \end{subfigure}
    ~
    \bfcaption{Error v. Number of Patterns. Note: y-axis is offset to better visualize difference.}   \label{fig:performance_vs_num_of_patterns}
    \trimfigurewhitespace
\end{figure}

\begin{figure}[h!]
	\captionsetup[subfigure]{justification=centering}
    \centering

    \begin{subfigure}[b]{0.48\textwidth}
        \centering       
        \includegraphics[width=\textwidth]{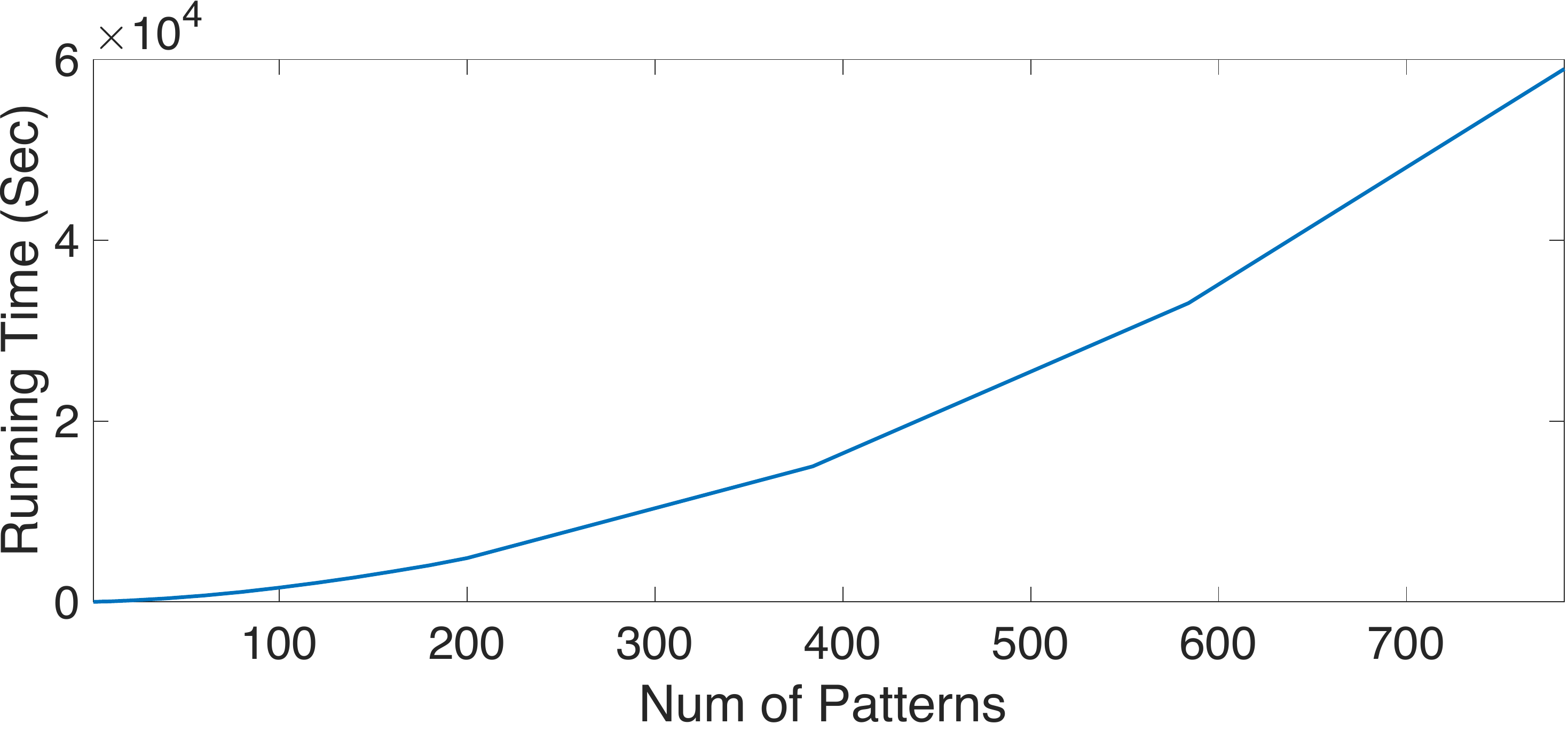}
 \bfcaption{Laserlight Run Time on Income Data}      \label{fig:laserlight_runningTimes_vs_NumOfPatterns}
\end{subfigure}
    ~
     \begin{subfigure}[b]{0.48\textwidth}
        \centering       
        \includegraphics[width=\textwidth]{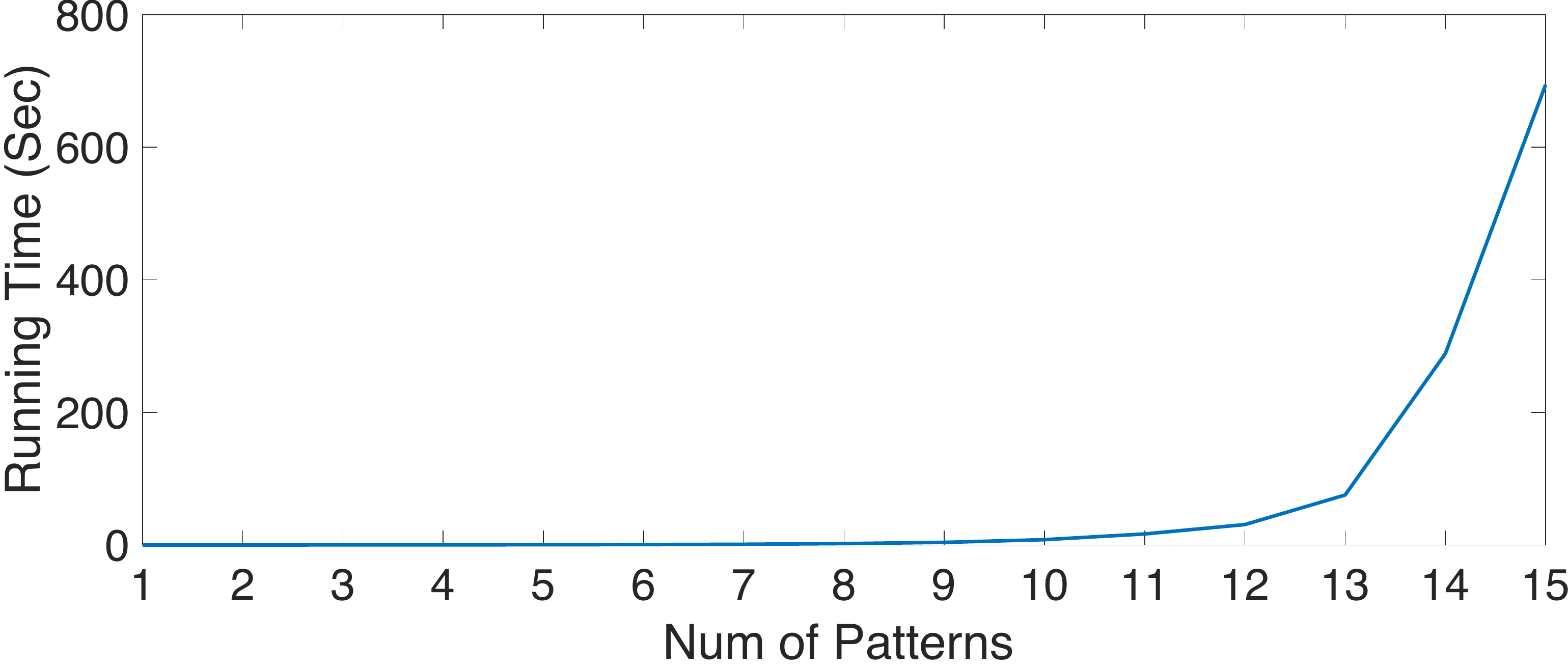}
 \bfcaption{MTV Run Time on Mushroom data}      \label{fig:mtv_runningTimes_vs_NumOfPatterns}
\end{subfigure}
    ~   
\bfcaption{Run Time of Laserlight and MTV}   \label{fig:runningTime_analysis}
\trimfigurewhitespace
\end{figure}

\subsubsection{Classical Laserlight and MTV}
\label{sec:classicallaserlightandmtv}

\tinysection{Establishing Baselines} 
To establish baselines, we evaluate \textit{Laserlight} and \textit{MTV} on their own data sets.
The results related to Error and run time are given in Figure~\ref{fig:performance_vs_num_of_patterns} and Figure~\ref{fig:runningTime_analysis} respectively.

In Figure~\ref{fig:performance_vs_num_of_patterns}, X-axis is the number of patterns and y-axis represents the Error measure of \textit{Laserlight} and \textit{MTV} in Figure~\ref{fig:Laserlight_Error_vs_NumOfPatterns} and~\ref{fig:MTV_Error_vs_NumOfPatterns} respectively.
We incorporate \textit{naive encoding} in Figure~\ref{fig:Laserlight_Error_vs_NumOfPatterns} as the reference method.
Since there are $783$ total number of features for \textit{Income} data set, the verbosity of \textit{naive encoding} will be $783$, which is shown as vertical dotted line in Figure~\ref{fig:Laserlight_Error_vs_NumOfPatterns}.
The \textit{Laserlight} Error of naive encoding is shown as the horizontal dotted line accordingly.
For \textit{Mushroom} data set (Figure~\ref{fig:MTV_Error_vs_NumOfPatterns}), the verbosity of its \textit{naive encoding} will be $96$.
However, \textit{MTV} quits with error message if it is requested to mine over $15$ patterns.
Hence for Figure~\ref{fig:MTV_Error_vs_NumOfPatterns}, the limit of x-axis is $15$ and we only show Error of \textit{naive encoding} as a reference line without marking out its verbosity. 
We observe in Figure~\ref{fig:Laserlight_Error_vs_NumOfPatterns} that \textit{naive encoding} outperforms \textit{Laserlight} when their verbosity is equal (i.e., $783$).
In addition, approximately after $100$ patterns, the slope of Error reduction becomes relatively flat.
Similar observations can be made from Figure~\ref{fig:MTV_Error_vs_NumOfPatterns}.
In Figure~\ref{fig:runningTime_analysis}, we observe that the running time increases exponentially with the number of patterns, for both \textit{Laserlight} and \textit{MTV}.

The take-aways from Figure~\ref{fig:performance_vs_num_of_patterns} and Figure~\ref{fig:runningTime_analysis} are that (1) \textit{naive encoding} is faster and more accurate than classical \textit{Laserlight} and \textit{MTV}; (2) the runtime increases superlinearly with the number of patterns mined from both \textit{Laserlight} and \textit{MTV}.

\tinysection{Anti-correlation and Dimentionality Reduction}
Recall in Section~\ref{sec:refiningnaivemixtureencodings} that \textit{Laserlight} is restricted to $100$ features.
For its own \textit{Income} data set, \textit{Laserlight} can be applied with its full set of $783$ features.
This is due to the prior knowledge that the $783$ features belong to $9$ groups.
In each group, features are mutually anti-correlated which can be reduced to a single feature.
Similarly, \textit{Mushroom} data set can be reduced from $95$ to $21$ features (See Table~\ref{table:extendeddatasummary}).

\subsubsection{Generalizing Laserlight and MTV}
\label{sec:generalizinglaserlightandmtv}

We generalize \textit{Laserlight} and \textit{MTV} on partitioned data by running them on each cluster.
We then combine Errors on all clusters by taking a weighted average, as described in Section~\ref{sec:generalizedinformationlossmeasures}.
Depending on how many patterns are mined from each cluster, \textit{Laserlight} and \textit{MTV} can be generalized into two types: (1) The number of patterns mined from each cluster is scaled to be equal to the verbosity of the \textit{naive encoding}; and (2) The total number of patterns mined from all clusters is fixed to a given number.
We name the first type \textit{Laserlight (MTV) Mixture Scaled}, which is comparable to \textit{naive mixture encoding}.
We name the second type \textit{Laserlight (MTV) Mixture Fixed}, which is comparable to the classical \textit{LaserLight (MTV)} algorithm.


The Error reduction of \textit{Laserlight} becomes relatively slow over $100$ patterns (See Figure~\ref{fig:Laserlight_Error_vs_NumOfPatterns}).    
Hence we configure \textit{Classical Laserlight} and \textit{Laserlight Mixture Fixed} to mine a total of $100$ patterns from the original data and partitioned data respectively, in order to avoid underestimation of \textit{Laserlight} performance.
We describe the strategy for distributing the $100$ patterns to partitions of the data sets in Appendix~\ref{appendix:Configuring_Regularized Laserlight_Mixture}.

\begin{figure}[h!]
	\captionsetup[subfigure]{justification=centering}
    \centering
    \begin{subfigure}[b]{0.48\textwidth}
        \centering       
        \includegraphics[width=\textwidth]{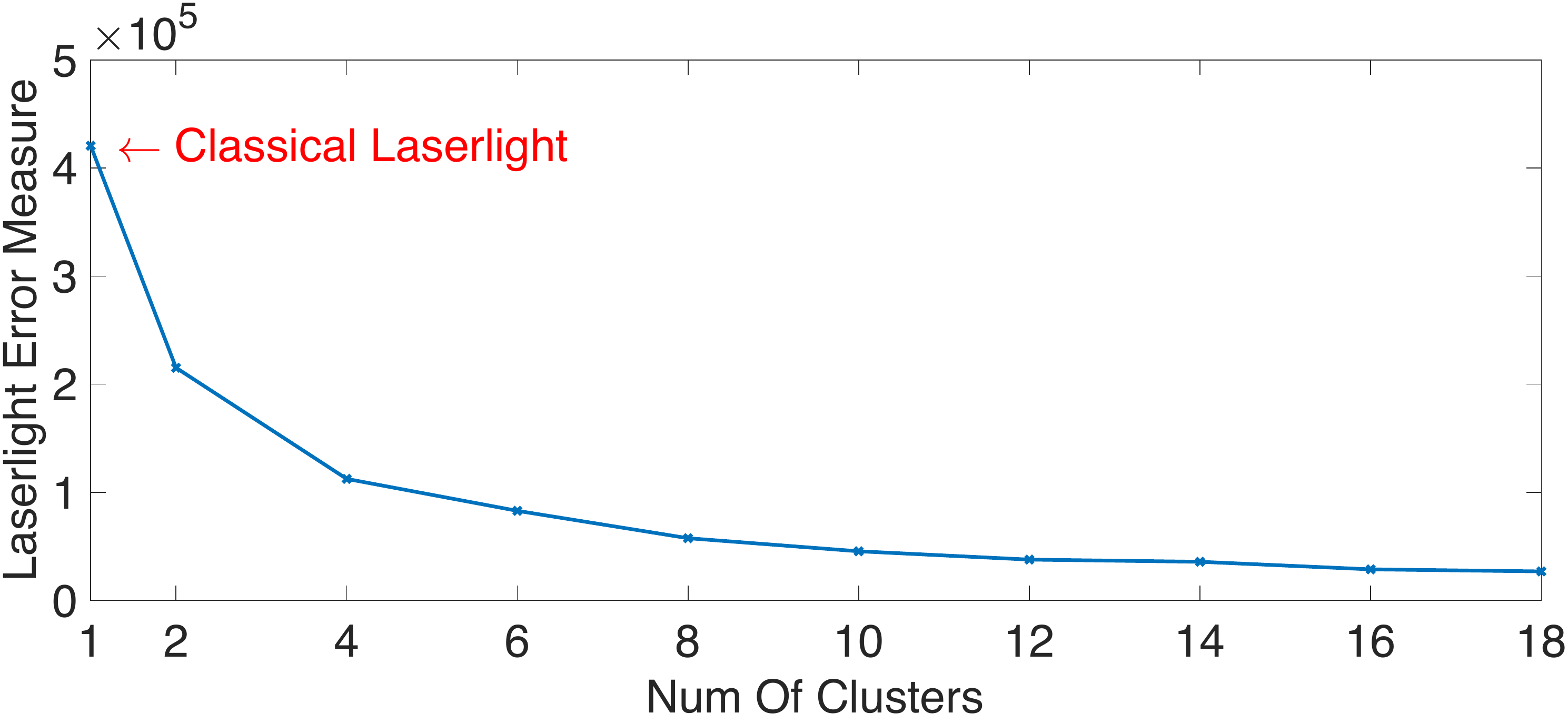}
 \bfcaption{Laserlight Error v. \# of Clusters on Income data}      \label{fig:RegularizedLaserlightMixture_Errors_vs_NumOfClusters}
\end{subfigure}
    ~
\begin{subfigure}[b]{0.48\textwidth}
  \centering       
  \includegraphics[width=\textwidth]{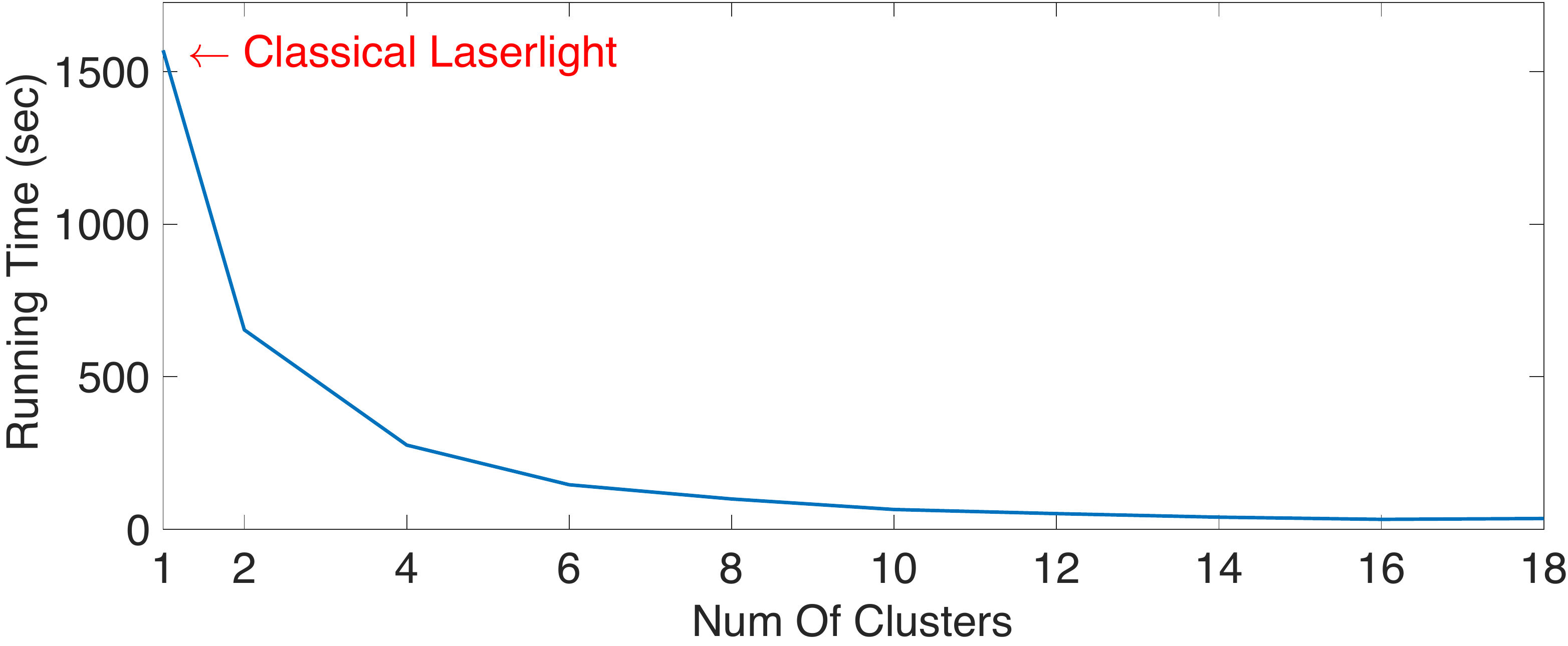}
 \bfcaption{Total Running time v. \# of Clusters on Income data}     
 \label{fig:RegularizedLaserlightMixture_vs_Laserlight_runningTimes_IncomeData}
\end{subfigure}
~
\bfcaption{Laserlight Mixture Fixed v. Classical}   
\label{fig:Regularized_Laserlight_Mixture_vs_Classical Laserlight}
\trimfigurewhitespace
\end{figure}

The experiment result is given in Figure~\ref{fig:Regularized_Laserlight_Mixture_vs_Classical Laserlight}.
Figure~\ref{fig:RegularizedLaserlightMixture_Errors_vs_NumOfClusters} and Figure~\ref{fig:RegularizedLaserlightMixture_vs_Laserlight_runningTimes_IncomeData} shows \textit{Laserlight Mixture Fixed} versus \textit{Classical Laserlight} in Error and run time respectively, when the number of data partitions (i.e., clusters) is gradually increased.
We observe an exponentially decreasing trend (i.e., improvement) in both \textit{Laserlight Error} and run time.
We omit the experiment results for \textit{MTV} as they give similar observations.

\tinysection{Take-away}
As the data is partitioned into more clusters, both runtime and Error of \textit{Laserlight (MTV) Mixture Fixed} exponentially improve.
This observation can be potentially generalized to other pattern mining based algorithms.

\subsubsection{Comparison with Naive Mixture Encoding}
\label{sec:Evaluating_Naive_Mixture_Encoding}

\begin{figure}[h!]
	\captionsetup[subfigure]{justification=centering}
    \centering
    \begin{subfigure}[b]{0.48\textwidth}
        \centering       
        \includegraphics[width=\textwidth]{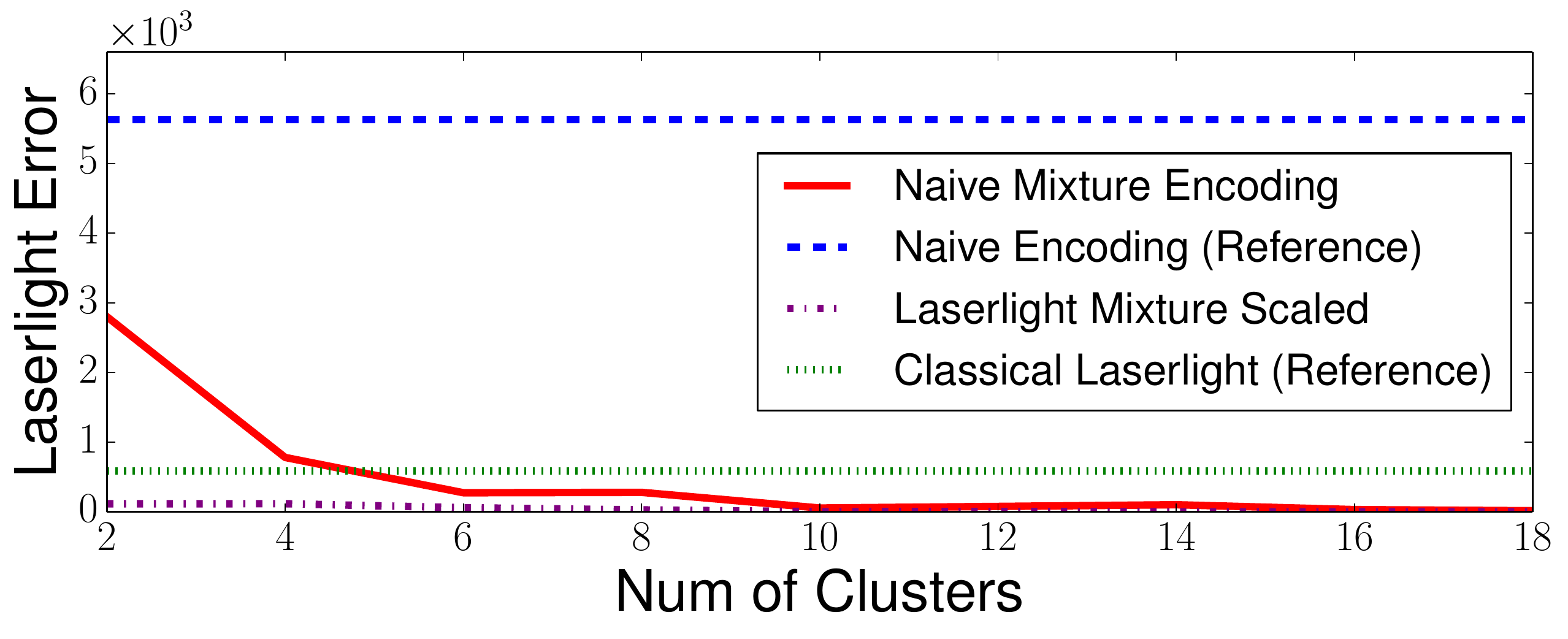}
 \bfcaption{Laserlight Error v. \# of Clusters on Mushroom data}      \label{fig:LaserlightMixture_Errors_vs_NumOfClusters}
\end{subfigure}
    ~
\begin{subfigure}[b]{0.48\textwidth}
  \centering       
  \includegraphics[width=\textwidth]{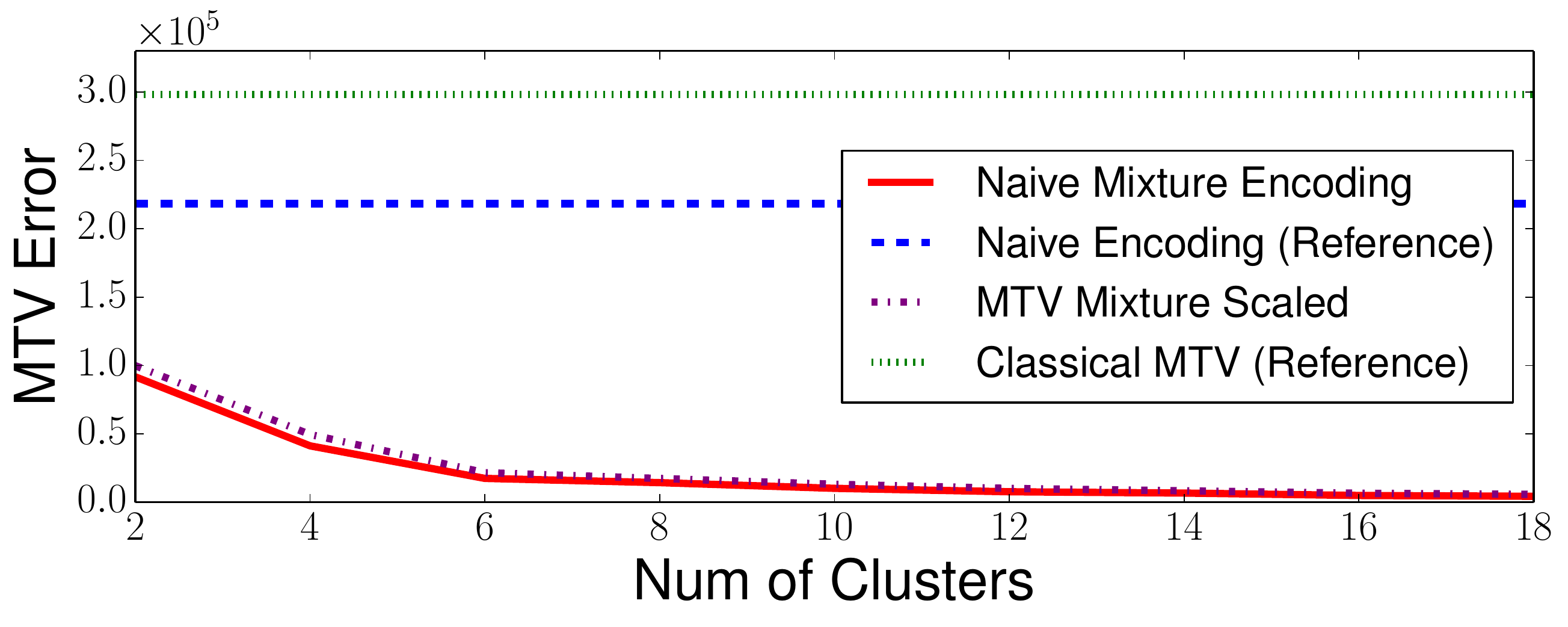}
 \bfcaption{MTV Error v. \# of Clusters on Mushroom data}     
 \label{fig:MTVMixture_vs_Laserlight_runningTimes_IncomeData}
\end{subfigure}
~
\bfcaption{Naive Mixture v. Laserlight/MTV Mixture}   
\label{fig:Naive Mixture Encoding_vs_Laserlight&MTV_Mixture}
\trimfigurewhitespace
\end{figure}

At last, we compare \textit{Laserlight (MTV) Mixture Scaled} with \textit{naive mixture encoding}.
Note that it is time-consuming for \textit{Laserlight} to mine the same number of patterns as \textit{naive encoding} on \textit{Income} data (See runtime analysis in Figure~\ref{fig:laserlight_runningTimes_vs_NumOfPatterns}), we choose \textit{Mushroom} data for \textit{Laserlight Mixture Scaled} instead.
The experiment results are given in Figure~\ref{fig:Naive Mixture Encoding_vs_Laserlight&MTV_Mixture}. 
The x-axes for all sub-figures in Figure~\ref{fig:Naive Mixture Encoding_vs_Laserlight&MTV_Mixture} represent the number of clusters and the y-axes stands for \textit{Laserlight} and \textit{MTV} Error respectively.
We incorporate baselines (i.e., \textit{naive encoding}, classical \textit{Laserlight} and \textit{MTV}) as reference lines in Figure~\ref{fig:LaserlightMixture_Errors_vs_NumOfClusters} and~\ref{fig:MTVMixture_vs_Laserlight_runningTimes_IncomeData} respectively.
We also experienced a limitation of $15$ patterns in configuring \textit{MTV}.
Hence the comparison between \textit{MTV Mixture Scaled} and {naive mixture encoding} is not strictly on equal footing as \textit{MTV Mixture Scaled} is not able to reach the same Total Verbosity as \textit{naive mixture encoding}.
Note that their difference in verbosity is mitigated by the fact that \textit{MTV} Error measure penalizes verbosity.

Figure~\ref{fig:LaserlightMixture_Errors_vs_NumOfClusters} shows that both \textit{naive mixture encoding} and \textit{Laserlight Mixture Scaled} have lower Error than their baselines.
In addition, \textit{Laserlight Mixture Scaled} has lower Error than \textit{naive mixture encoding} when the number of clusters is less than $4$ and they become close after $6$ clusters.
In other words, \textit{Laserlight} is more accurate on lightly partitioned data. 
As the data is further partitioned, clusters become `easier' to summarize, and \textit{naive encoding} becomes more similar to \textit{Laserlight}.
Figure~\ref{fig:MTVMixture_vs_Laserlight_runningTimes_IncomeData} shows that \textit{naive mixture encoding} marginally outperforms \textit{Laserlight Mixture Scaled}.


\tinysection{Take-away}
\textit{Naive mixture encoding} is faster and has similar (lower) Error than \textit{Laserlight (MTV) Mixture Scaled}.

\section{Related Work}
\label{sec:backgroundandrelatedwork}
\subsection{Workload Analysis}
Existing approaches related to workload analysis are frequently aimed at specific tasks like query recommendation~\cite{5693465, Giacometti:2009:QRO:1651291.1651306,Khoussainova:2010:SCA:1880172.1880175,yang2009recommending,aligon2014similarity}, performance optimization~\cite{aouiche2006clustering,bruno2001stholes}, outlier detection~\cite{kamra2008detecting} or visual analysis~\cite{makiyama2015text}. 

\tinysection{Query Recommendation}
This task aims to track historical querying behavior and generating query recommendations.
Related approaches~\cite{5693465,Khoussainova:2010:SCA:1880172.1880175} flatten a query \textit{abstract syntax tree} as a bag of \textit{fragments}~\cite{5693465} or \textit{snippets}~\cite{Khoussainova:2010:SCA:1880172.1880175} and adopt \textit{feature vector} representation of queries.
User profiles are then built from the query log by grouping and summarizing feature vectors by user in order to make personalized recommendation. 
Under OLAP systems, profiles are also built for workloads of similar OLAP sessions~\cite{aligon2014similarity}.

\tinysection{Performance Optimization}
Index selection~\cite{chaudhuri1997efficient,finkelstein1988physical} and materialized view selection~\cite{Agrawal:2000:ASM:645926.671701,aouiche2006clustering,bruno2001stholes} are typical performance optimization tasks.
The configuration search space is usually large, but can be reduced with appropriate summaries.

\tinysection{Outlier Detection}
Kamra \textit{et al.}~\cite{kamra2008detecting} aim at detecting anomalous behavior of queries in the log by summarizing query logs into profiles
of normal user behavior.

\tinysection{Visual Analysis}
Makiyama \textit{et al.}~\cite{makiyama2015text} provide a set of visualizations that facilitate further workload analysis on Sloan Digital Sky Survey (SDSS) dataset.
QueryScope~\cite{hu2008queryscope} aims at finding better tuning opportunities by helping human experts to identify patterns shared among queries. 

In these approaches, queries are commonly encoded as feature vectors or bit-maps where a bit array is mapped to a list of features with $1$ in a position if the corresponding feature appears in the query and $0$ otherwise.
Workloads under the bit-map encoding must then be compressed before they can be efficiently queried or visualized for analysis. 

\subsection{Workload Compression Schemes}
\tinysection{Run-length Encoding}
\textit{Run-length encoding (RLE)} is a loss-less compression scheme commonly used in \textit{Inverted Index Compression}~\cite{Witten96managinggigabytes:,Zobel:2006:IFT:1132956.1132959} and \textit{Column-Oriented Compression}~\cite{Abadi:2006:ICE:1142473.1142548}.
RLE-based compression algorithms include but not limited to: Byte-aligned Bitmap Code (BBC) used in Oracle systems~\cite{Antoshenkov1996}, Word-aligned Hybrid (WAH)~\cite{1029710} and many others~\cite{Moffat94compressionand,Amer-Yahia:2000:OQC:645926.671874,515586}.
In general, RLE-based methods focus on column-wise compression and requires additional heavyweight inference on frequencies of cross-column (i.e., row-wise) patterns used for workload analysis.

\tinysection{Lempel-Ziv Encoding}
Lempel-Ziv~\cite{1055714,Ziv:2006:CIS:2263333.2269400} is the loss-less compression algorithm used by gzip.
It takes variable sized patterns (row-wise in our case) and replaces them with fixed length codes, in contrast to Huffman encoding~\cite{huffman1952method}. 
Lempel-Ziv encoding does not require knowledge about pattern frequencies in advance and builds the pattern dictionary dynamically. 
There are many other similar schemes for compressing files represented as sequential bit-maps, e.g.~\cite{skibinski2007fast}.

\tinysection{Dictionary Encoding}
\textit{Dictionary encoding} is a more general form of Lempel-Ziv.
It has the advantage that patterns with frequencies stored in the dictionary can be interpreted as workloads statistics useful for analysis.
In this paper, we extend dictionary encoding and focus on using a dictionary to infer frequencies of patterns not in it.
Mampaey \textit{et al.} proposed \textit{MTV} algorithm~\cite{Mampaey:2012:SDS:2382577.2382580} that finds the dictionary (of given size) having the lowest \textit{Bayesian Information Criterion(BIC)} score.
Lower BIC score indicates better accuracy on frequency inference.
Gebaly \textit{et al.} proposed \textit{Laserlight} algorithm~\cite{ElGebaly:2014:IIE:2735461.2735467} that builds a pattern dictionary for a slightly different goal.
The quality of the dictionary depends on whether it can correctly infer the truth-value of some binary feature.

\tinysection{Generative Models}
A generative model is a lossy compressed representation of the original log.
Typical generative models are \textit{probabilistic topic models}~\cite{Blei:2012:PTM:2133806.2133826,Wang:2009:MSU:1667583.1667675} and \textit{noisy-channel} model~\cite{Knight:2002:SBS:604203.604207}.
Generative models can infer pattern frequencies but they lack a model-independent measure for efficiently evaluating overall inference accuracy.

\tinysection{Low-Rank Matrix Decomposition}
Low-rank matrix decomposition~\cite{eckart1936approximation}, e.g. Principal Component Analysis (PCA) and Non-negative matrix factorization (NMF)~\cite{lee1999learning}, offers lossy data compression.
But the resulting matrices after decomposition are not suited for inferring workload statistics.

\section{Conclusion}
\label{sec:conclusion}
In this paper, we introduced the problem of log compression and defined a family of pattern based log encodings. 
We precisely characterized the information content of logs and offered three principled and one practical measures of encoding quality: Verbosity, Ambiguity, Deviation and \errorname. 
To reduce the search space of pattern based encodings, we introduced the idea of partitioning logs into separate components, which induces the family of pattern mixture as well as its simplified form: naive mixture encodings. 
Finally, we experimentally showed that naive mixture encodings are more informative and can be constructed more efficiently than encodings constructed from state-of-the-art pattern based summarization techniques. 

\tinysection{Future Work} 
The use of mixture models for summarization has potential implications for work on pattern mining; As we show, existing techniques can be substantially improved.  
We also expect that making accurate correlated feature counting efficient will enable a range of more powerful database tuning and intrusion detection systems.

\vspace{\textheight minus \textheight} 

\pagebreak

\bibliographystyle{acm}
\bibliography{paper,oliver} 

\appendix
\section{NOMENCLATURE}
 \begin{tabular}{|c c|} 
 \hline
 \textbf{Symbol} & \textbf{Meaning}  \\ [0.5ex] 
 \hline\hline
 $f$ & Feature\\ 
 \hline
 $\vec{b}$ & Pattern \\
 \hline
  $\vec{b'} \subseteq \vec{b}$ & $\vec{b'}$ is contained in $\vec{b}$\\
 \hline  
 $\vec{q}$ & Query\\
 \hline
 $L$ & Log, a bag of queries\\
  \hline
 $Q$ & Query randomly drawn from $L$\\
 \hline 
  $p(Q\;|\;L)$ & Query distribution of $L$\\
 \hline
 $p(Q\supseteq\vec{b}\;|\;L)$ & Marginal probability of $Q\supseteq\vec b$\\
 \hline
  $\encoding_L[\pattern]$ & same as  $p(Q\supseteq\vec{b}\;|\;L)$\\
 \hline
   $corr\_rank(\vec{b})$ & Feature-correlation score\\ 
 \hline 
  $\encoding_{max}$ & Mapping from all patterns to marginals\\
 \hline 
  $\encoding$ & Encoding, partial mapping $\encoding \subseteq \encoding_{max}$\\
  \hline 
  $domain(\cdot)$ & Domain of mapping\\
 \hline
  $\rho$ & An arbitrary query distribution\\
 \hline
  $\Omega_\encoding$ & Space of $\rho$ constrained by $\encoding$\\
 \hline
   $\encoding\leq_{\Omega}\encoding'$ & $\Omega_{\encoding}\subseteq\Omega_{\encoding'}$\\ 
   \hline
   $\mathcal{P}_{\encoding}$ & A random $\rho$ drawn from $\Omega_\encoding$\\
  \hline 
  $\rho^*$ & Same as $p(Q\;|\;L)$, $\rho^*\in\Omega_\encoding$\\
 \hline 
    $\mathcal{H}(\cdot)$ & Entropy of Distribution\\
 \hline 
  $\overline{\rho}_\encoding$ & Representative distribution of $\Omega_\encoding$\\
 \hline 
 $\text{d}(\encoding)$ & Deviation\\
 \hline 
  $\text{I}(\encoding)$ & Ambiguity\\
 \hline 
  $e(\encoding)$ & \errorname \\
 \hline
   $|\encoding|$ & $|domain(\encoding)|$, Verbosity of encoding\\
   \hline
  $\rho\ll\rho'$ & $\rho$ is absolutely continuous w.r.t $\rho'$\\
 \hline 
 $\mathcal{D}_{KL}(\rho||\rho')$ & K-L Divergence from $\rho'$ to $\rho$.\\
 \hline
\end{tabular}

\section{Proof of Proposition~\ref{proposition:losslesssummary}}
\label{appendix:losslesssummary}
Denote by $\mathds{1}^n = \{0,1\}^n$ the space of possible 0-1 vectors of size $n$, and define an encoding $\bar \encoding_{\vec{q}}$ with patterns:
$$domain(\overline{\encoding}_{\vec{q}})=\comprehension{(x_1+b_1,\ldots,x_n+b_n)}{(b_1,\ldots,b_n)\in\mathds{1}^n}$$
We will show that $\overline{\encoding}_{\vec{q}}\subseteq \encoding_{max}$ contains sufficient information to compute $p_0 = p(X_1=x_1,\ldots,X_n=x_n)$ through several steps.
First, we define a new pair of marginal probabilities $p_1\tuple{b_1} = p(X_1 \geq x_1 + b_1,X_2=x_2,\ldots,X_n=x_n)$.
$x_1$ is integral, so $p_0 = p_1\tuple{0} - p_1\tuple{1}$.
Generalizing, we can define:
\begin{multline*}
p_k\tuple{b_1, \ldots, b_k} = p(X_1 \geq x_1 + b_1,\; \ldots,\; X_k \geq x_k + b_k, \\
  X_{k+1} = x_{k+1},\; \ldots,\; X_n=x_n)
\end{multline*}
Again, $x_k$ being integral gives us that:
\begin{multline*}
p_{k-1}\tuple{b_1, \ldots, b_{k-1}} = p_{k}\tuple{b_1, \ldots, b_{k-1}, 0} \\
  -p_{k}\tuple{b_1, \ldots, b_{k-1}, 1}
\end{multline*}
Finally, when $k = n$, the probability $p_n\tuple{b_1, \ldots, b_n}$ is the marginal probability $p(Q\supseteq\vec{b}\;|\;L)$ of a pattern $\vec{b}=(x_1+b_1, \ldots, x_n+b_n)$, which by definition is offered by $\overline{\encoding}_{\vec{q}}$ for any $(b_1, \ldots, b_n) \in \mathds{1}^n$.
The resulting encoding $\overline{\encoding}=\bigcup_{\vec{q}\in L}\overline{\encoding}_{\vec{q}}$ identifies the distribution $p(Q\;|\;L)$, which we refer to as \emph{lossless} encoding. Clearly any encoding that extends $\overline{\encoding}$ (including $\encoding_{max}$) is lossless.

\section{Sampling From Space of Distributions}
\label{appendix:sampling}
Here we describe how we sample a random distribution $\rho$ from the space $\Omega_\encoding$ of probability distributions.
\subsection{Preliminary Sampling}
To sample a random distribution $\rho$ which assigns a probability value to each element in vector space $\mathbb{B}^n$. 
The naive way is to treat $\rho$ as a random multi-dimensional vector $(\rho(\vec{q}_1),\ldots,\rho(\vec{q}_{|\mathbb{B}^n|}))$ that sum up to 1.
\begin{algorithm}
\caption{Sampling}
\label{alg:sampling}
\begin{algorithmic}[1]
\Procedure{TwoStepSampling}{}
\label{procedure:twostepsampling}
\State Step 1:
\For{each $\vec{v}\in \mathbb{B}^m\wedge\mathcal{C}_{\vec{v}}\neq\emptyset$}
\State $V \gets V\; \bigcup \;\vec{v}$
\EndFor
\State $class\_p \gets $\Call{UniRandDistribProb}{V,$1$}
\State Step 2:
\For{each $\vec{v}\in V$}
\State $\rho \gets \rho\; \bigcup \; $\Call{UniRandDistribProb}{$\mathcal{C}_{\vec{v}}$,$class\_p(\vec{v})$}
\EndFor
\State \Return $\rho$
\EndProcedure

\State

\Procedure{UniRandDistribProb}{Set S, double prob}
\For{each element $e \in S$}
\State $p(e) \gets UniformRandNum(range=[0,1])$
\EndFor
\For{each element $e \in S$}
\State $p(e) \gets prob\times p(e)\div\mysuml_{e}p(e)$
\EndFor
\State \Return $p$
\EndProcedure
\end{algorithmic}
\end{algorithm}
However, $|\mathbb{B}^n|$ is exponentially large (i.e., $2^n$) and we reduce the number of elements in $\rho$ by grouping them (i.e., $\vec{q}_1\ldots,\vec{q}_{|\mathbb{N}^n|}$) into equivalence classes.

\tinysection{Encoding-equivalent Classes}
The basic idea for grouping is based on containment relationship between query $\vec q_i$ and patterns $\vec{b}\in \encoding$ in the encoding $\encoding$. 
More precisely, if $\vec{q}_i\supseteq\vec{b}$, it indicates that the assignment $\rho(\vec{q}_i)$ on $i$th dimension is constrained by marginal $\encoding[\vec{b}]$ (See Section~\ref{sec:lossysummaries}).
As a result, if queries $\vec{q}_i,\vec{q}_j$ share the same containment relationship with pattern $\vec{b}$, assignments $\rho(\vec{q}_i),\rho(\vec{q}_j)$ on $i$th and $j$th dimension make no difference for satisfying the constraint of pattern $\vec{b}$.
We thus define \textit{pattern-equivalence} as $$\vec{q}_i\equiv_{\vec{b}}\vec{q}_j\Leftrightarrow BI(\vec{q}_i,\vec{b})=BI(\vec{q}_j,\vec{b})$$ 
$BI(\vec{q}_i,\vec{b})$ is the Binary Indicator function satisfying $BI(\vec{q}_i,\vec{b})=1\equiv\vec{q}_i\supseteq\vec{b}$. 
Queries are \textit{encoding-equivalent} $\vec{q}_i\equiv_{\encoding}\vec{q}_j$ if they are pattern-equivalent for all patterns in the encoding. 
Numbering patterns in the encoding as $\vec{b}_1,\ldots,\vec{b}_m$, any binary vector $\vec{v}\in\mathbb{B}^m$ maps to an equivalence class $\mathcal{C}_{\vec{v}}=\comprehension{\vec{q}}{(BI(\vec{q},\vec{b}_1),\ldots,BI(\vec{q},\vec{b}_m))=\vec{v}\wedge\vec{q}\in\mathbb{N}^n}$.
Though the number of non-empty equivalent classes may grow as large as $\mathcal{O}(2^m)$, it is much smaller than $2^n$ in most cases and sampling a random distribution $\rho$ can be divided into two steps as shown in line~\ref{procedure:twostepsampling} of algorithm~\ref{alg:sampling}. 
Note that $class\_p$ in the algorithm, which is produced by the first step, is a randomly sampled distribution over all non-empty equivalence classes. The second step redistributes probabilities randomly assigned to each equivalence class to its class members in an unbiased way.

\subsection{Incorporating Constraints}
So far we are creating random samples from an unconstrained space of distributions. 
To make sure $\rho$ produced by the two-step sampling fall within space $\Omega_\encoding$, the probabilities distributed over equivalence classes (denoted as $class\_p$) must obey the linear equality constraints derived from the encoding $\encoding$.
Denote the space of candidate $class\_p$ as $U$ and the subspace allowed by the encoding as $U_\encoding\subseteq U$, one naive solution is to reject $class\_p\notin U_\encoding$. 
However, the subspace $U_\encoding$ constrained under linear equality constraints is equivalent to an intersection of \textit{hyperplanes} in the full space $U$. 
The volume of $U_\encoding$ is thus infinitely small comparing to that of $U$, such that any random sample $class\_p\in U$ will \textit{almost never} fall within $U_\encoding$.
To make sampling feasible, we do not reject a sample $class\_p\in U$ but \textit{project} it onto the hyperplane of $U_\encoding$ by finding its \textit{closest} (Euclidean distance) counterpart $class\_p'\in U_\encoding$:
$$class\_p'=\argminl_{class\_p'\in U_\encoding}||class\_p'-class\_p||_2$$
Finding the projection point $class\_p'$ of $class\_p$ can be achieved by linear programming.

\section{Algorithm Configurations}
\label{appendix:experimentsettingsforpatternbasedalgorithms}
Here we give detailed description on our selected state-of-the-art pattern based summarizers (i.e., \textit{Laserlight} and \textit{MTV}) and also specify how we configured them in experiments discussed in Section~\ref{sec:motivatepatternmixturesummaries}. 

\tinysection{Common Configuration}
We set up both algorithms to mine $15$ patterns from target clusters.
This is because, empirically we found that \textit{MTV} quits with error message over $15$ patterns.
For fair comparison, we set the same number of patterns for \textit{Laserlight}. 

\subsection{Laserlight Algorithm}
\label{appendix:Laserlight}
\tinysection{Description}
\textit{Laserlight} algorithm is proposed in~\cite{ElGebaly:2014:IIE:2735461.2735467} for summarizing multi-dimension data (i.e., $D=(X_1,\ldots,X_n)$) augmented by a binary attribute $A$. 
The goal is to search for a set of patterns (i.e., encoding) from the data $D$ that provide maximum information for predicting augmented attribute $A$, which is a sub-problem of summarizing the joint distribution $p(D,A)$. 
Another algorithm \textit{Flashlight} is also proposed in the same paper but we omit it in our experiment due its inferior scalability.
The implementation of \textit{Laserlight} has been incorporated into PostgreSQL 9.1 and the source code is only available upon request.

\tinysection{Experiment Settings}
Due to the restriction on the maximum number of data dimension by the PostgreSQL implementation of $Laserlight$, we project the distribution $p(Q\;|\;L)$ onto a limited set of $100$ features.
The selection criteria is based on feature entropy or variability. 
More precisely, regarding the existence of $i$th feature as random binary variable $X_i$, features are ranked by entropy $\entropy(X_i)$. 
The feature with highest entropy $\entropy(X_i)$ is chosen as the augmented attribute $A$.
The algorithm heuristically selects a limited set of samples from the space of candidate patterns, from which the pattern that is most informative is selected to be added to the encoding.
Note that when we applied $Laserlight$ in our experiments, we set the number of samples to be $16$, which is suggested in ~\cite{ElGebaly:2014:IIE:2735461.2735467} based on its own data sets.

\subsection{MTV Algorithm}
\tinysection{Description}
\textit{MTV} algorithm is proposed in~\cite{Mampaey:2012:SDS:2382577.2382580} for summarizing multi-dimensional data with binary attributes. 
The goal is to mine a succinct set of patterns (i.e., encoding) that convey the most important information (See the paper for definition). 
The implementation of this algorithm can be obtained at 
\href{http://adrem.ua.ac.be/succinctsummary}{http://adrem.ua.ac.be/succinctsummary}.

\tinysection{Experiment Settings}
\textit{MTV} requires to set the minimum support threshold for patterns. That is, patterns with marginal less than the threshold will be ignored, in order to reduce the search space of candidate patterns.
We set the minimum support threshold to be $0.05$ in our experiments such that any pattern that is contained in more than $5\%$ of queries will be considered as candidate.

\subsection{Configuring Laserlight Mixture Fixed}
\label{appendix:Configuring_Regularized Laserlight_Mixture}
Given a data partitioning and fixed total number of patterns to mine from all clusters, in order to determine the number of patterns mined from each cluster, we need to assign weights $\sum_i w_i=1$ for each cluster $i$. 
\errorname $e(\encoding_L)$ of the naive encoding $\encoding_L$ for a cluster reflects its `easiness' for pattern mining and the intuition is that $e(\encoding_L)=0$ indicates there is no need for additional pattern mining.
\errorname $e(\encoding_L)$ is affected by the number of features $n$ ever occur in the cluster.
Consider a toy data with only two feature vectors $\vec v_1=(0,0)$ and $\vec v_2=(1,1)$. Appending $\vec v_1$ with new features of value $0$ and $\vec v_2$ of value $1$ will increase $e(\encoding_L)$ but not necessarily the number of patterns needed for accurately summarizing the data.
Hence we normalize $e(\encoding_L)$, dividing it by the number of features $n$, which gives us $w_i=\frac{e(\encoding_L)}{n}$.
In addition, since the \textit{generalized measure} for \textit{Laserlight} and \textit{MTV} gives weight to each cluster proportional to its number of distinct data instances $m$, we also adjust $w_i$ and multiply them with the number of distance data instances.
The final weight assignment becomes $w_i \propto \frac{m}{n}e(\encoding_L)$.

\section{Interpreting Naive Mixture Encoding}
\label{appendix:naivemixturesummaryvisualization}
Due to sensitive information contained in the US bank data set, we only provide visualization on PocketData.

The visualization of PocketData is based on its naive mixture encoding under $8$ clusters\footnote{The number of clusters is chosen for convenience of visualization.}. 
The result is given in Figure~\ref{fig:visualizepocketdatabyitsnaivemixturesummary}.
There are 5 sub-figures with each representing a naive encoding for one cluster. 
Note that we use shading to represent the magnitude of marginals and features with marginal too small will be invisible and omitted.
Question mark `$?$' is the placeholder for constants.
Three clusters from the eight are not shown in the figure: One cluster is too messy (i.e., further sub-clustering is needed) and two clusters gives similar visualization to Figure~\ref{fig:cluster1} and~\ref{fig:cluster5}.
The caption of each sub-figures expresses our understanding on the task that queries in the cluster are performing, by visualizing the corresponding naive encodings.
For simplicity, we will also omit features of \texttt{SELECT} category if they are neither participating in \texttt{WHERE} clause nor intuitively related to other features in \texttt{SELECT}. 
\begin{figure}[h!]
 \centering
\begin{subfigure}{\columnwidth}
  {\small
    \begin{tabular}{r|p{60mm}}
    \textbf{SELECT} & 
            {\texttt{conversation\_id}}, {\texttt{participants\_type}},
        {\texttt{first\_name}},
        {\texttt{chat\_id}},
        \texttt{blocked},
        \texttt{active}\\ \hline
    \textbf{FROM} &
        \texttt{{\texttt{conversation\_participants\_view}}}\\ \hline
    \textbf{WHERE} &
        \texttt{(chat\_id!=?)} $\wedge$
        {\texttt{(conversation\_id=?)}} $\wedge$
        \texttt{(active=1)}   
    \end{tabular}
  }
  \bfcaption{Check the person who is active in specific conversation and not participating in specified chat.}
  \label{fig:cluster1}
\end{subfigure}\\[2mm]

\begin{subfigure}{\columnwidth}
  {\small
    \begin{tabular}{r|p{60mm}}
    \textbf{SELECT} & 
            {\texttt{status}}, 
        {\texttt{timestamp}},
        {\texttt{expiration\_timestamp}},
        {\texttt{sms\_raw\_sender}},
        \texttt{message\_id},
        \texttt{text}\\ \hline
    \textbf{FROM} &
        \texttt{{\texttt{conversations}}},
        \textcolor{mid-gray}{\texttt{{\texttt{message\_notifications\_view}}}},
        \texttt{{\texttt{messages\_view}}}\\ \hline
       \textbf{ORDER BY} &
       \texttt{{\texttt{Descend on timestamp}}}
       \\ \hline
        \textbf{Limit} &
       \texttt{{\texttt{500}}}
       \\ \hline
    \textbf{WHERE} &      
        \textcolor{mid-gray}{\texttt{(expiration\_timestamp>?)}} $\wedge$
        \texttt{(status!=5)} $\wedge$
        {\texttt{(conversation\_id=?)}} $\wedge$        
{\texttt{(conversations.conversation\_id=conversation\_id)}}   
    \end{tabular}
  }
  \bfcaption{Check sender information for most recent SMS messages that participate in given conversation.}
  \label{fig:cluster2}
\end{subfigure}\\[2mm]

\begin{subfigure}{\columnwidth}
  {\small
    \begin{tabular}{r|p{60mm}}
    \textbf{SELECT} & 
            {\texttt{status}}, 
        {\texttt{timestamp}},
        {\texttt{conversation\_id}},
        {\texttt{chat\_watermark}},
        \texttt{message\_id},
        \texttt{sms\_type}\\ \hline
    \textbf{FROM} &
        \texttt{{\texttt{conversations}}},
        \texttt{{\texttt{message\_notifications\_view}}}\\ \hline
    \textbf{WHERE} &
        \texttt{(conversation\_status!=1)} $\wedge$
         \texttt{(conversation\_pending\_leave!=1)} $\wedge$
         \texttt{(conversation\_notification\_level!=10)} $\wedge$
         \texttt{(timestamp>1355...)} $\wedge$
         \texttt{(timestamp>chat\_watermark)} $\wedge$
        {\texttt{(conversation\_id=?)}} $\wedge$        
{\texttt{(conversations.conversation\_id=conversation\_id)}}   
    \end{tabular}
  }
  \bfcaption{Check recent messages in conversations of specific type.}
  \label{fig:cluster3}
\end{subfigure}\\[2mm]

\begin{subfigure}{\columnwidth}
  {\small
    \begin{tabular}{r|p{60mm}}
    \textbf{SELECT} & 
            {\texttt{suggestion\_type}}, {\texttt{name}},
        {\texttt{chat\_id}}\\ \hline
    \textbf{FROM} &
        \texttt{{\texttt{suggested\_contacts}}}\\ \hline
       \textbf{Limit} &
        \textcolor{light-gray}{\texttt{{\texttt{10}}}}\\ \hline
        \textbf{Order By} &
        \textcolor{light-gray}{\texttt{{\texttt{Ascend on upper(name)}}}}\\ \hline
    \textbf{WHERE} &
        \texttt{(chat\_id!=?)} $\wedge$
        {\texttt{(name!=?)}}
    \end{tabular}
  }
  \bfcaption{Suggest contacts that avoid certain names and chat.}
  \label{fig:cluster4}
\end{subfigure}\\[2mm]

\begin{subfigure}{\columnwidth}
  {\small
    \begin{tabular}{r|p{60mm}}
    \textbf{SELECT} & 
            \textcolor{mid-gray}{{\texttt{sms\_type}}},  \textcolor{mid-gray}{{\texttt{timestamp}}},
         \textcolor{mid-gray}{{\texttt{\_id}}}\\ \hline
    \textbf{FROM} &
        \texttt{{\texttt{messages}}}\\ \hline
    \textbf{WHERE} &
        \texttt{(sms\_type=1)} $\wedge$
        {\texttt{(status=4)}} $\wedge$
         {\texttt{(transport\_type=3)}} $\wedge$
         \textcolor{mid-gray}{{\texttt{(timestamp>=?)}}}
    \end{tabular}
  }
  \bfcaption{Check messages under type/status conditions}
  \label{fig:cluster5}
\end{subfigure}\\[2mm]

\bfcaption{\textbf{Visualize PocketData by its naive mixture encoding}}
\label{fig:visualizepocketdatabyitsnaivemixturesummary}
\trimfigurewhitespace
\end{figure}

\end{document}